\documentclass[12pt,a4paper]{article}
\usepackage{setspace}

\usepackage[top=1in, bottom=1in, left=1in, right=1in]{geometry}

\usepackage{amssymb}
\usepackage[utf8]{inputenc} % allow utf-8 input
\usepackage[T1]{fontenc}    % use 8-bit T1 fonts
\usepackage{hyperref}       % hyperlinks
\usepackage{url}            % simple URL typesetting
\usepackage{booktabs}       % professional-quality tables
\usepackage{amsfonts}       % blackboard math symbols
\usepackage{nicefrac}       % compact symbols for 1/2, etc.
\usepackage{microtype}      % microtypography
\usepackage{lipsum}
\usepackage{graphicx}
\graphicspath{ {./images/} }
\usepackage{amsmath}
\usepackage{bm}% bold math
\usepackage{color}
\usepackage{hyperref}

\pdfoutput=1
\newcommand{\hl}{ }

\hypersetup{
    colorlinks=true,
    linkcolor=blue,
    filecolor=blue,      %magenta
    urlcolor=blue,  %cyan
    citecolor = blue,
    pdfpagemode=FullScreen,
    }

\title{Machine Learning for High-entropy Alloys: Progress, Challenges and Opportunities \footnote{\footnotesize{This manuscript has been co-authored by UT-Battelle, LLC, under contract DE-AC05-00OR22725 with the US Department of Energy (DOE). The US government retains and the publisher, by accepting the article for publication, acknowledges that the US government retains a nonexclusive, paid-up, irrevocable, worldwide license to publish or reproduce the published form of this manuscript, or allow others to do so, for US government purposes. DOE will provide public access to these results of federally sponsored research in accordance with the DOE Public Access Plan (http://energy.gov/downloads/doe-public-access-plan).}}}

\author{
 \textbf{Xianglin Liu} \\
  \\
  Peng Cheng Laboratory\\
  Oak Ridge National Laboratory\\
\and 
\textbf{Jiaxin Zhang} \\ 
  \\
Oak Ridge National Laboratory\\
  \and 
  \textbf{Zongrui Pei} \\ 
  \\
New York University\\
Oak Ridge National Laboratory\\
\texttt{zp2137@nyu.edu}\\
}

\begin{document}

\maketitle
\begin{abstract}
%{\color{red}
High-entropy alloys (HEAs) have attracted extensive interest due to their exceptional mechanical properties and the vast compositional space for new HEAs. However, understanding their novel physical mechanisms and then using these mechanisms to design new HEAs are confronted with their high-dimensional chemical complexity, which presents unique challenges to (i) the theoretical modeling that needs accurate atomic interactions for atomistic simulations and (ii) constructing reliable macro-scale models for high-throughput screening of vast amounts of candidate alloys. Machine learning (ML) sheds light on these problems with its capability to represent extremely complex relations. This review highlights the success and promising future of utilizing ML to overcome these challenges. We first introduce the basics of ML algorithms and application scenarios. We then summarize the state-of-the-art ML models describing atomic interactions and atomistic simulations of thermodynamic and mechanical properties. Special attention is paid to phase predictions, planar-defect calculations, and plastic deformation simulations. Next, we review ML models for macro-scale properties, such as lattice structures, phase formations, and mechanical properties. Examples of machine-learned phase-formation rules and order parameters are used to illustrate the workflow. Finally, we discuss the remaining challenges and present an outlook of research directions, including uncertainty quantification and ML-guided inverse materials design.

\end{abstract}

\tableofcontents
\doublespacing

\section{Introduction}
\subsection{High-entropy alloys }
The conventional strategy of alloying is to add a small amount of property-enhancing elements into one primary metal. Humans have used this technique for thousands of years, exemplified by some of the most widely used materials in human history, such as bronzes, steels, and aluminum alloys. However, the discovery of the single-phase multicomponent high-entropy alloys (HEAs) in 2004 \cite{ADEM:ADEM200300567, CANTOR2004213} introduced a different alloying strategy. While conventional alloys are comprised of one base element with a few additional elements of much lower concentrations, in HEAs, multiple-principal elements are mixed to obtain a single-phase random solid solution. These new classes of materials are named HEAs \cite{ADEM:ADEM200300567, EGeorge_Nature} due to their supposedly enhanced configurational entropy, which was believed to be the cause of their stabilization. However, the role of configurational entropy is not always as important as originally assumed \cite{EGeorge_Nature, OTTO20132628, MA201590}, therefore names such as multicomponent alloys, multi-principal-element alloys, compositionally-complex alloys, or complex concentrated alloys have also been widely used \cite{CANTOR2021100754}. This new strategy of mixing multiple principal elements opens the door to exploring the vast uncharted space beyond the corners of chemical compositional diagrams. HEAs have become one of the most exciting research directions in materials science, and HEAs of unprecedented performance have been constantly emerging, as demonstrated in Fig.~\ref{fig:Opportunity} b. Since there are a number of excellent review papers focused on HEAs \cite{tsai2014high,zhang2014microstructures,miracle2017critical,george2019high,george2020high,ikeda2019ab,ma2019tailoring}, we have no intention to restate the details therein, which is also not the main focus of this work.

Several key factors contribute to the enormous interest in HEAs. The first one is their {\it exceptional mechanical properties}, such as the exceptional high-temperature properties \cite{SENKOV2011698}, enhanced strength, ductility and toughness at cryogenic temperatures \cite{Gludovatz1153, NiCoCr_cryo}, and most importantly the excellent combination of strength and ductility \cite{CANTOR2004213, NatureRaabe, OxygenHEA, PKLiawNC2019, Yang933, ORNL_HEAs_2021, RitchieNatureComm, DamageToleranceGeorge, Fueaat8712, BetaPrimePhaseNature2020,doi:10.1126/science.abj8114}. Some HEAs demonstrate the unusual capability of overcoming the strength-ductility tradeoff \cite{NatureRaabe, Yang933, ORNL_HEAs_2021}. Another reason is that HEAs also provide an attractive platform to study deformation mechanisms related to {\it chemical order and disorder}. The different occupations of lattice sites by the multiple principal elements give rise to various degrees of randomness and order, characterized by short-range order (SRO) \cite{SRO_HEA_2020,chen2021direct,Ding8919} and long-range order (LRO) parameters \cite{PhysRevB.91.224204, widom_2018}, which have a significant impact on the physical quantities crucial for plastic deformation, such as dislocation core structures and generalized stacking fault energies \cite{ZHANG2019424, ZENG20191}\hl{\cite{ LIU2019107955}}. Tuning the degree of order and disorder can improve the mechanical properties of HEAs \cite{NatureCommSai, HU2021100854, SRO_HEA_2020, chen2021direct, Ding8919} \hl{\cite{pei2020statistics}}. 
A third reason is their {\it diverse and heterogeneous microstructures}, such as the polymorphology of stacking faults \hl{\cite{PhysRevLett.126.255502}},  twin architectures \cite{RitchieNatureComm}, decoupling of between Shockley partials and stacking faults \hl{\cite{pei2021decoupling}}, nanoscale phases \cite{NatureRaabe}, ultrafine-grained lamella structures \cite{PKLiawNC2019}, and nano-precipitates \cite{Yang933, BetaPrimePhaseNature2020, ORNL_HEAs_2021, https://doi.org/10.1002/advs.202100870}. While these microstructures can present in conventional alloys, the large number of elements in HEAs bring more degrees of freedom to tune them. The high tunability is beneficial to enhancing the mechanical properties of single-phase random HEAs. Last but not least, the barely explored non-equiatomic concentrations in the {\it vast, high-dimensional compositional space} of HEAs, demonstrate the huge potential for materials design. 

\subsection{Challenges in modeling high-entropy alloys}
While the increased principal elements bring huge opportunities to alloy design, they also introduce significant challenges to the theoretical modeling and simulations. The first one is the difficulty in constructing empirical atomic interaction models, mainly due to the large number of chemical interactions involved. For instance, for an $n$-component system, the number of $m$-site interactions proportional to 
\begin{equation*}
C(n,m)=\frac{n!}{m!(n-m)!}=\frac{n(n-1)\cdots (n-m+1)}{m(m-1)\cdots 1}.
\end{equation*}
For a five-element system, only considering the pair interactions within the first four coordination shells already give $4\times C(5,2) = 40$ different interactions. Considering the higher-order interactions would substantially further increase the number of interactions, making the traditional cluster expansion method \cite{PhysRev.81.988, SANCHEZ1984334, vandeWalle2002, widom_2018} and embedded atom method (EAM) \cite{VARVENNE2016164} prone to overfitting. The second challenge occurs in the first-principles simulations. It is well-known that density functional theory (DFT) simulations are generally computationally expensive. However, they are more prohibitive for HEAs due to the necessity of using large supercells (up to more than 1000 atoms) to represent the non-stoichiometric compositions, complex order-disorder behaviors, microstructures, and extended dislocation core structures \cite{SMITH2016352} \hl{\cite{LIU2019107955}}. The computational cost is even more demanding for first-principles thermodynamics simulations, where many configurations must be evaluated, each calculated from the expensive DFT. For instance, for a 250-atom CuZn supercell, the first-principles Monte-Carlo simulation requires the self-consistent (SCF) calculations of 600,000 DFT energies \cite{PhysRevB.93.024203}, which can easily consume a million CPU core-hours. Finally, the traditional trial-and-error method of materials design is too laborious for HEAs, therefore substitute models are highly desirable to guide the efficient exploration of the vast compositional space, such as phase formation rules \cite{PhaseFormation2014}\hl{\cite{ pei2020machine}}, strengthening model  \cite{VARVENNE2016164, WU2016108} and ductility criteria \hl{\cite{pei2019machine}}\cite{MAK2021104389,HU2021116800}. Nevertheless, these empirical rules are challenging to formulate for HEAs due to the large number of chemical species involved.

\subsection{Merits of machine learning for high-entropy alloys}
\begin{figure} [ht!]
    \centering
    \includegraphics[width=0.8 \linewidth]{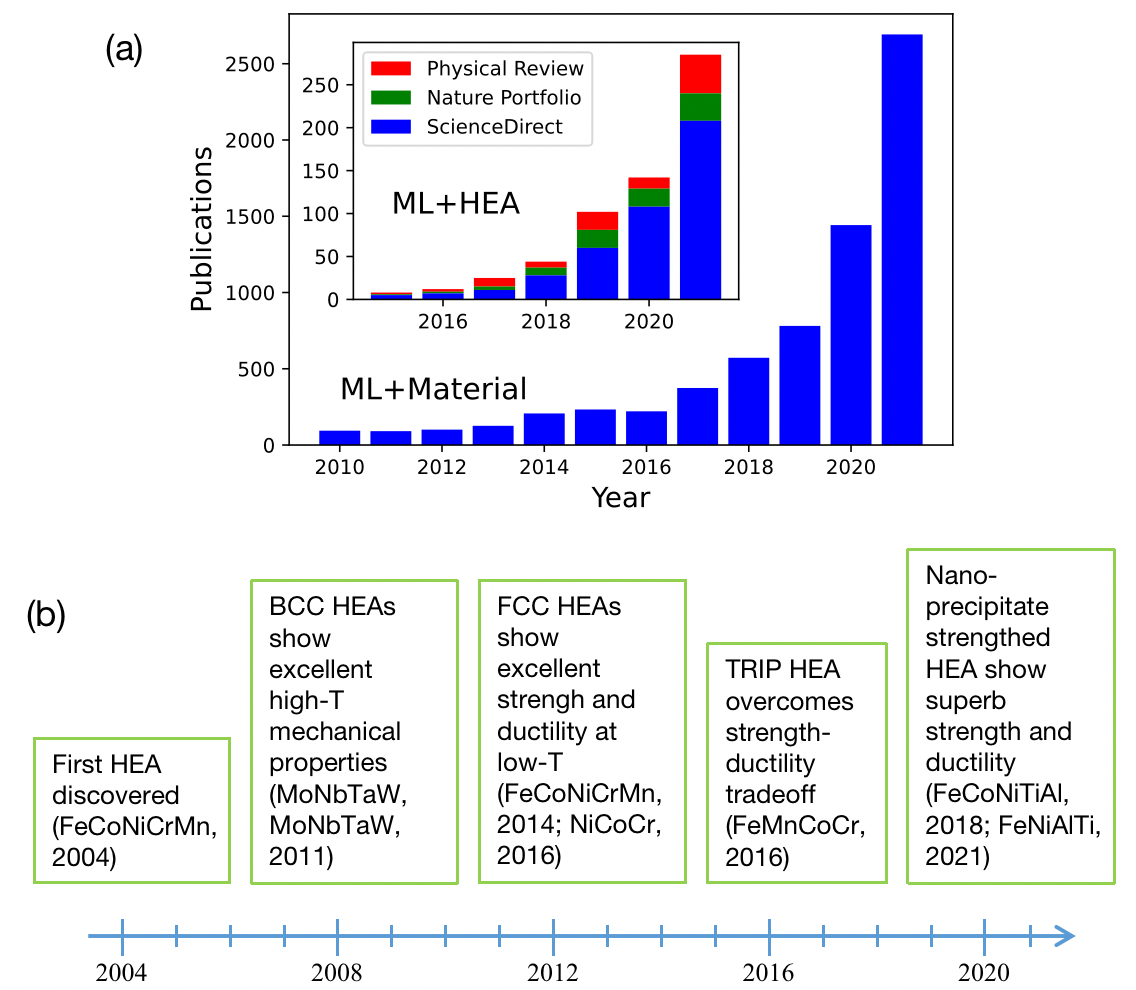}
    \caption{The progresses in high-entropy alloys and machine learning. (a) Annual publications retrieved with the keywords ``machine learning material'' and ``machine learning high-entropy alloys'' (inset plot) from ScienceDirect (blue) within the subject area of materials science, Nature Portfolio journals (green), and Physical Review journals (red). Search date: July 26, 2022. (b) Time table of a few representative high-performance HEAs.}
    \label{fig:Opportunity}
\end{figure}

The past decade has witnessed the explosive rise of machine learning (ML) and its sub-domain deep learning \cite{DeepLearning}. This revolution has not only fundamentally changed several domains of computer science, such as computer vision and natural language processing, but also a plethora of scientific fields ranging from particle signal detection \cite{Particle_ML}, protein structure prediction \cite{AlphaFold}, medical image analysis \cite{LITJENS201760}, to the survey of the universe \cite{NatureUniverse}. The success of ML lies in its capability to describe complex patterns, and these complex ML models can be systematically optimized by learning from the data. This capability is further powered by readily available computing resources, efficient algorithms, and big data collected from experiment or computation. These features make ML a promising tool to address the aforementioned challenges confronted by the theoretical modeling of HEAs. The exponential growth of relevant publications reflects this trend (see Fig.~\ref{fig:Opportunity} a). In principle, given a sufficiently large high-quality data, the complex atomic interactions can be well-captured by ML through the standard procedures of training, validation, and testing.

%Specifically, ML can capture complex atomic interactions with its nonlinear hierarchical representation capability. 
%A large number of features from the various atomic interactions can be well-treated through the standard ML training, validation, and testing procedures, given sufficient high-quality data.

Moreover, the calculation speed of ML models is typically much faster than DFT and close to empirical potentials \cite{GNNFF_NPJ_2021, Hart_NPJ_2019, Hart_NPJ_2021, Ong_JPC_2020}. This high efficiency allows researchers to simulate materials with more than millions of atoms near DFT accuracy \cite{10.5555/3433701.3433707, Nature_2021_Silicon}, way beyond the limit of conventional DFT. 
%Such capability is crucial for investigating the defects and structures at the nanoscale \cite{Nature_2021_Silicon}. 
Such capability is particularly important for understanding the extraordinary mechanical properties in HEAs, originating from the complex ordering, defects, and microstructures  \cite{Asta_NC_MTP_2021, Ong_NPJ_SFE} that demand accurate nanoscale simulations. Besides simulations, ML also provides a powerful tool to build prediction models for material properties. By learning from the data, ML can automatically identify features or latent variables essential for achieving the desired physical properties. These complex, nonlinear structure-property relationships are challenging to formulate by human observation. Identifying them allows researchers to explore the enormous design space of HEAs more efficiently.
The above merits of ML has ushered in a new data-driven paradigm to the research of HEAs \cite{DURODOLA2021100797, ML_alloys}, as demonstrated in its success in the atomistic simulations \hl{\cite{ZHANG2020108247, LIU2021110135,yin2021neural}}, physical property prediction \hl{\cite{pei2020machine}}\cite{ NatureCommCCAHard}, and materials design \cite{ha2021evidence, DDJohnsonML2021Nature,rao2022machine}.

\subsection{Uniqueness and structure of this review}

This review focuses on the application of ML for high-entropy alloys (HEAs). The progress, challenges, and opportunities of this specific topic are within the scope of the review. While several excellent reviews have already been available on the broader topic of ML for materials science \cite{ ML_material_science, NPJ_MaterialsGenome, NPJ_RecentAdvanceML, doi:10.1063/1.5126336, doi:10.1146/annurev-matsci-070218-010015, Nature_MLPotential, ML_alloys, DURODOLA2021100797}, there is a lack of comprehensive reviews for the exciting progress and huge potential of ML for HEAs. For instance, Drodola reviewed the applications of ML for alloys in a recent publication \cite{DURODOLA2021100797}, which focused on using artificial neural networks for alloy design, processing, and characterization. Another review on ML for alloys was presented by Hart \textit{et al.} \cite{ML_alloys}, where they summarized the current state of machine-learning-driven alloy research. Both review papers summarize ML applications for alloys in general. Still, ML's unique challenges and opportunities for HEAs are not thoroughly discussed. Given the extensive attention that HEAs have attracted, such a timely, focused review is needed. To be specific,

\begin{itemize}
    \item {\it We present a comprehensive review on the interdisciplinary topic of ML for HEAs, discuss the basic methods and models heuristically, and summarize the important applications to illustrate the advantages of ML for HEAs;}    
    \item {\it We identify challenges in the theoretical modeling of HEAs, and illustrate how ML can help tackle these problems} from both {\it microscopic} and {\it macroscopic} perspectives;
    \item {\it We compare the strengths and shortcomings of different ML methods for HEAs, summarize the remaining challenges of ML methods, and discuss the opportunities accompanied by these challenges in the future.}
\end{itemize}

This review is intended to pinpoint the opportunities in ML for HEAs and provide guidance to researchers interested in exploring this new field. To this end, we will introduce the ML methods, survey their applications in multicomponent alloys, and elucidate the remaining challenges. This review is organized as follows: First, an introduction to ML for HEAs is given, including application identification, data generation, common algorithms, model training, and results analysis. Next, popular ML potentials for atomistic simulations are discussed. The following three sections focus on the recent progress in applying ML to HEAs, including atomistic simulations, property predictions, and materials design. Finally, the challenges and opportunities of ML for HEAs will be discussed in the outlook section.

\section{How to utilize machine learning for high-entropy alloys?}
\begin{figure} [h]
    \centering
    \includegraphics[width=0.7 \linewidth]{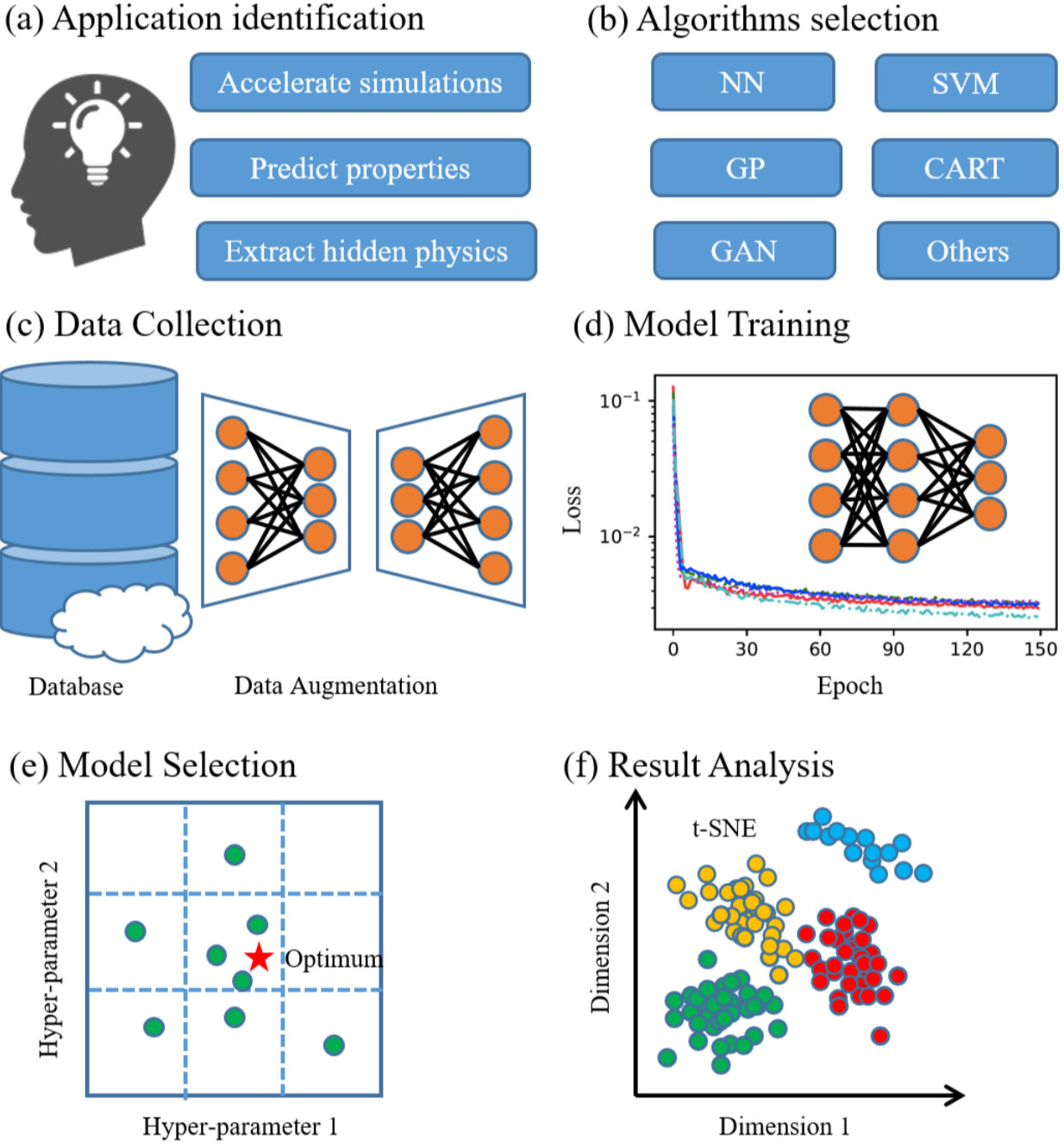}
    \caption{The critical steps in a typical ML application for HEAs. (a) Application identification. Typical applications include accelerating simulations, predicting physical properties, and extracting hidden physics. (b) Selection of appropriate machine learning algorithms, such as neural networks (NN), support vector machine (SVM), Gaussian process (GP), classification and regression tree (CART), and generative adversarial network (GAN). (c) Data collection. The data can be obtained from a database and augmented with techniques such as GAN. (d) Model training. A typical example is stochastic gradient descent to optimize the parameters in neural networks. (e) Model selection. Grid search and Bayesian optimization can be used. (f) Result analysis. The model parameters can be analyzed with dimension-reduction techniques.
    }
    \label{fig:steps}
\end{figure}

%\subsection{General workflow}
The ML model's general workflow is illustrated in Fig.~\ref{fig:steps}. While ML is generally a versatile and powerful tool, one should keep in mind that it only shines for a certain subset of problems. Specifically, ML is most effective for problems with {\it abundant high-dimensional data} and {\it well-defined objectives}. Therefore, the first step for utilizing ML for HEAs is {\it application identification} \cite{ML_Domain_NC2020}, i.e., confirming that the problem can be converted to an ML one. After that, appropriate ML algorithms should be chosen to solve the specific ML problem. The data are collected and generally split into three subsets, i.e., (i) the training set used for model training, (ii) the validation set for model selection, and (iii) the test set for model performance evaluation. Finally, a well-trained ML model should capture the underlying physics. Good-performance models render it possible to extract the knowledge by inspecting their parameters directly or through dimension reduction techniques \cite{zhang2019learning}.

%\textit{For the synergy of ML and materials science, the three most important factors to be considered are data, algorithms, and applications. While the importance and data and algorithms in machine learning are well-known, identifying appropriate applications that can fully take advantage of the potential of ML is also crucial. In the following we discuss the three important factors in detail.}
%
%\begin{itemize}
%\item how to get data
%\item how to build model: how to use ML? how to represent a chemical configuration. 
%\item how ML could be useful?
%
%\item overcome problems of ML: small-data, undersampling, multi-dimension, too large design space
%\end{itemize}
\subsection{Application identification}
One of the most common ML applications for HEAs is to build {\it surrogate models}. Surrogate models are used to approximately model a set of input-output data when the actual relationship is unknown or difficult to deduce. 
Currently, the applications of these surrogate models can be roughly grouped into (i) atomistic simulations on the atomic scale and (ii) physical property predictions on the macro scale. Here we refer to the corresponding ML models as atomic interaction models (AIMs) and bulk property models (BPMs). The applications are compared in Tab.~\ref{tab:t1}. Specifically, AIM and BPM tackle materials science problems from different time and length scales. AIM follows the spirit of first-principles methods. However, instead of subatomic electrons and nuclei, as in DFT, the input feature of AIM is the arrangement of different atoms. In the language of the renormalization group, the electron degree of freedom has been  ``integrated out''. Compared to BPM, one major benefit of AIM is its high transferability. Well-trained AIM can be employed to evaluate many different physical quantities, such as the formation energy, elastic constants, stacking fault energies, and phonon spectra \cite{Hart_NPJ_2019, Hart_NPJ_2021}. By comparison, the advantage of BPM is that it directly predicts the target physical properties without the involvement of expensive simulations. Another key difference between AIM and BPM is the fidelity requirement: AIMs generally have a higher requirement on the accuracy of the model prediction than BPMs. For instance, a nonphysical force prediction in one step in molecular dynamics may lead all the following simulations to a completely wrong trajectory. Therefore, 
it is important to tailor physical information into AIMs, such as the materials' underlying translational, rotational, and permutational symmetry.  

\begin{table}[!ht] 
%\footnotesize
\centering
\caption{Comparison of ML surrogate models for atomistic simulations and bulk property predictions.}
\label{tab:t1}
\begin{tabular} {p{0.2\textwidth}p{0.35\textwidth}p{0.35\textwidth}} %{cccc }
\hline
\hline
%\hline
%\hline 
 & Atomistic Simulation & Bulk Property Prediction \\ 
\hline
\hline
Role of ML & Providing empirical potentials to accelerate simulations & Predicting measurable or calculable physical properties
 \\ \hline
Targets & Atomic energies, forces, and stresses
& Phase formation, crystal structure, elastic constants \\ \hline
Data source & First-principles calculations & First-principles calculations, experimental data \\ \hline
Data generation	& Computationally expensive (hundreds to millions of CPU hours)	&Experimentally expensive ($>$ hundreds of experiments)  \\ \hline
Descriptors & Invariant representation of atomic environment & Properties of chemical elements, relevant physical properties \\ \hline
Models/Algorithms & Physical Descriptors + ML 
& Traditional ML algorithms (e.g., support vector machine, decision trees, Gaussian process) and neural networks \\
\hline
%Time-to-solution & Typically not expensive& Typically not expensive \\ \hline
Model inference & 	Needed for each atom at every simulation step &	Needed for exploring new materials \\ \hline \hline
%Model fidelity &	High &	Medium \\ \hline \hline
\end{tabular}
Acronyms: ML- machine learning; CPU-central processing unit.
\end{table}

In addition to building surrogate models, ML can also be employed to draw physical insights from extensive data. To this end, reducing data dimensions and extracting important features are critical since the physical rules are more evident in small dimensions. For some ML models, such as decision trees and shallow neural networks \cite{KAUFMANN2020178, ML_Phase_sensitivity}, essential features can be identified by inspecting the weight parameters of the models. Unsupervised learning is also a widely used technique to extract hidden physics. Unsupervised learning does not require labor-intensive data labeling, which is one of its unique advantages. Instead, it explores the patterns and structures in the unlabeled dataset. A typical example of unsupervised learning is clustering, which divides the data into groups according to their similarity in the feature space. Clustering can be used as a dimension reduction technique to classify a large number of features into a few groups. With the grouped features, the relation between the desired properties and a few labeled data can be established straightforwardly \cite{YingZhangNC2020}.
Another excellent example is the variational autoencoder (VAE) \cite{kingma2013auto}. It can significantly reduce the data size and simultaneously keep the essential information in the latent space with only very few dimensions, usually two, for better visualization. Assisted by this advantage of VAE, Yin, Pei, and Gao proposed a VAE-based order parameter to describe the phase transitions in HEAs \hl{\cite{yin2021neural}}. The new order parameter and VAE model are based on many chemical configurations generated using Monte-Carlo simulations, informed by the interaction parameters also obtained from ML models. The order parameters for phase transitions are one of the central topics in condensed matter physics and materials science, so this also demonstrates the incredible power of ML contributions to physics.

\subsection{Representative machine-learning algorithms}

% update

Different ML algorithms have different strengths and shortcomings. Therefore, it is crucial to choose the appropriate one according to the task objective and dataset size. For example, when the dataset size is small, a simple linear regression should perform better than a deep neural network model due to the well-known bias-variance tradeoff \cite{MEHTA20191}. There are already many excellent introductions to ML algorithms \cite{MEHTA20191,DeepLearning}, so here we only briefly describe a few representative ones.

\textbf{Neural network (NN)}:
Neural networks, also referred to as artificial neural networks (ANNs), are the backbone of deep learning algorithms \cite{lecun2015deep,goodfellow2016deep}. A neural network is comprised of multiple function layers, with each layer composed of multiple neurons. A  neuron receives inputs from its previous layer, multiplies them (dot-product) with its weight parameters, adds the bias term, and passes the result to the activation function to produce the output. One simple example of a neural network is the multilayer perceptron, which is a feed-forward network comprised of a few fully connected layers. Some other frequently used neural network structures include the convolutional neural network (CNN) \cite{lecun1995convolutional}, recurrent neural network (RNN) \cite{hochreiter1997long}, and transformers \cite{vaswani2017attention}. NNs are typically optimized with stochastic gradient descent (SGD) algorithms, in which the backpropagation method is commonly employed to update the model parameters, aiming to minimize the loss function for given data samples. The loss function, by its name, measures the deviation from the desired distribution. For supervised learning, the loss function typically measures the difference between the model prediction and the true target values. Choosing an appropriate loss function is very important for ML. For regression tasks, the quadratic loss function is a common choice, while the cross entropy loss is widely used for classification tasks.

%The NN models are widely used to classify the phases and crystal structures of multicomponent alloys. Compared to the linear output layer in regression models, a logistic or softmax function is used in the output layer for classification. To avoid vanishing or exploding gradients in DNN, the ReLU activation function is preferred, and techniques such as batch normalization and regularization are typically applied. Cross entropy is commonly used as the loss function, the optimization of which can be achieved iteratively via stochastic gradient descent (SGD), with the weights updated by back-propagation.

\textbf{Support vector machine (SVM)}: SVM is a widely used algorithm for separating different groups of data \cite{cortes1995support}. The principle of SVM is to find a hyperplane that maximizes the separation between the two groups. The hyperplane decides which group a data point belongs to. In linear SVM, the decision plane $\mathbf{w}^T \mathbf{x} + b=0$ that maximizes the separation is given by minimizing $||\mathbf{w}||$ under the constraint $y_i(\mathbf{w}^T \mathbf{x}_i+b)\geq 1$, for all data $(\mathbf{x}_i, y_i)$, where $y_i\in \{-1,1\}$. As a result, the decision plane only depends on the closest data points unaffected by the others. Other than the above ``hard margin'' version, SVM can also be easily extended to cases where the data are not linearly separable. A regularization term is usually used to penalize the margin violation, which is called soft-margin SVM. Moreover, the separation boundary in SVM can also be extended to nonlinear cases by employing the so-called kernel method, which amounts to implicitly projecting the data to a high-dimensional feature space. One common choice for the kernel is the radial basis function (RBF), which takes the form of a Gaussian function and has infinite feature space. Finally, multi-class SVM can be formulated by decomposing the problem into multiple binary classification problems other than binary classification.

\textbf{Gaussian process (GP)}: GP is based on Bayes' theorem, i.e., the posterior is proportional to the product of prior and likelihood \cite{rasmussen2003gaussian}. In GP regression, each data $x$ is assumed to follow one Gaussian distribution, and nearby data points $x$ and $x'$ are related smoothly by the covariance function (kernel). Due to the multivariate Gaussian distribution assumption, the posterior can be easily expressed in terms of the covariance matrix of the data points. Therefore, the training process essentially evaluates the covariance matrix and its inverse, from which the probability distribution of model parameters can be obtained. During the inference stage, the prediction is obtained by averaging the output of all possible models. The most attractive feature of GP is that it can estimate the probability distribution of the model parameters, which makes uncertainty quantification straightforward \cite{rasmussen2010gaussian}. On the other hand, the Gaussian process is generally computationally expensive for a large dataset ($>$ 10,000 data points), in which case some sparsity technique is needed to reduce the cost by choosing a sub-sample of the entire dataset \cite{snoek2012practical}. Other than regression, GP can also be used for classification tasks by appending a logistic or softmax function. However, introducing a non-Gaussian function makes the model difficult to solve, and methods such as the Laplace approximation are needed to avoid the computationally intensive Monte-Carlo sampling.

\textbf{Classification and regression tree (CART)}
A decision tree is an intuitive and easy-to-interpret method \cite{timofeev2004classification}. It starts from a root node, and criteria are chosen for the features to split the data into subsets, as represented by the descendant nodes. This process continues until the termination condition is reached. The splitting criteria are based on a randomness measure for the subset. One widely used measure is the Gini impurity, defined as $G=1-\sum_{i} p_i^2$, where $p_i$ is the ratio of class $i$ in the node. In terms of the Gini impurity, the cost function for splitting the data into $n$ subsets, i.e. Gini index, is given by $L=\sum_i\frac{m_{i}}{m} G_i$
% \begin{align*}
% L=\sum_i\frac{m_{i}}{m} G_i,
% \end{align*}
where $m_i$ is the number of data in the $i$-th subset, and $m=\sum_i
 m_i$. A single DT is a relatively weak predictor, and the performance can be improved by combining a group of predictors, each trained with a data subset from bootstrap aggregating (bagging). When each decision tree uses a random set of features for splitting, such methods are called random forest (RF), a widely used ensemble method \cite{zhou2019ensemble}. A stronger predictor can also be obtained by sequentially improving the performance; such an ensemble method is known as boosting \cite{chen2016xgboost}. Typical examples include AdaBoost, and Gradient Boosting \cite{MEHTA20191}.

%\subsubsection{Other important algorithms}

\subsection{Data collection}
\subsubsection{Data for atomistic simulations}
The training datasets are generally obtained via DFT calculations for constructing atomic-interaction surrogate models. A high-quality training dataset for ML models should be \textit{homogeneous, sufficiently large}, and \textit{representative}. By homogeneous, we mean the DFT energies are calculated using the same method, with the same set of parameters, so that the energy difference between data can truly reflect the impact of atomic configuration. Important DFT parameters include the type of exchange-correlation functional, the energy cutoff of plane waves, the mesh grid of k-points, etc. A sufficiently large dataset is necessary for the model to capture the underlying physics accurately. Finally, a representative dataset means the sampled data points are evenly distributed in the configuration space. This representativity is crucial for an accurate surrogate model since ML algorithms are usually much more reliable doing interpolation than extrapolation.

The linear-scaling DFT methods can be utilized to improve the efficiency of DFT calculations. Linear-scaling DFT reduces the computational complexity of DFT from $O(N^3)$ to $O(N)$ by making use of the nearsightedness principle \cite{Prodan11635}; therefore, it is highly advantageous for simulating HEAs, where supercells with more than 100 atoms are commonly needed. Different linear-scaling DFT methods have been proposed and implemented \cite{RevModPhys.71.1085}. For example, the locally self-consistent multiple-scattering (LSMS) method \cite{PhysRevLett.75.2867, osti_1420087} is employed by Liu {\it et al.} \hl{\cite{2019arXiv190602889L, liu2019chemical, ZHANG2020108247, LIU2021110135}} to accelerate the calculation of HEA dataset. The linear scaling in LSMS is achieved by restricting the quantum scattering of electrons within the so-called local interaction zone, but the long-range electrostatic interactions are still evaluated everywhere.

\subsubsection{Data for physical-property models}
There are various sources for the training data of ML models that predict bulk physical properties. These data can be experimental data or simulation data. Aided by powerful supercomputers and efficient algorithms, the volume of simulation data grows quickly. In particular, the well-structured simulation data are obtained with controllable and transparent conditions, which are easily repeatable and findable. These advantages render them attractive as fuel to power ML models.

{\it Experimental database}
Pauling File is an excellent experiment database  \cite{villars2004pauling}. The Pauling File project was launched in 1995 with the aim to create tools for scientists working with inorganic compounds, with a particular focus on materials design.
Another important one is the Materials Experiment and Analysis Database (MEAD) \cite{Soedarmadji2019}. In addition, some websites also provide collected experimental data for pure elements, such as periodictable.com. Pei {\it et al.} have used the data to train several ML models with good performance \hl{\cite{pei2019machine,pei2020machine,pei2021mechanisms}}.

{\it Computational database}
There are multiple well-acknowledged DFT databases for materials worldwide. Some well-known examples include the Materials Project of Lawrence Berkeley National Laboratory, Open Quantum Mechanics Database (OQMD) of Northwestern University by Wolverton {\it et al.} \cite{saal2013materials}, and AFlow by Curtarolo {\it et al.} at Duke University \cite{curtarolo2012aflow} in the U.S.A. In Europe, NOMAD is a representative one. Draxl and Scheffler reviewed some recent progress in big-data-driven materials science, particularly their effort to build the so-called FAIR data infrastructure \cite{Draxl2020}. Another European DFT database is the Automated Interactive Infrastructure and Database for Computational Science (AiiDA) which can be accessed by its interactive platform, Materials cloud \cite{talirz2020materials}.
Himanen {\it et al.} provided a good summary of the databases \cite{Himanen2019}.

There are also published data sources specifically for HEAs. For example, in 2018, two databases of HEAs were compiled by  Senkov and Miracle {\it et al.}, with one focus on the mechanical properties of 122 refractory HEAs \cite{COUZINIE20181622}, and the other one on 370 HEAs \cite{GORSSE20182664}. These databases were updated in 2020 by Borg {\it et al.} \cite{ScientificDataHEA2020}, with 1545 records of mechanical properties and phases from 265 articles being compiled. Gao {\it et al.} also published 1252 data of multicomponent solid solutions and intermetallics \cite{gao2017thermodynamics}, which was employed by Pei {\it et al.} to train an ML model for phase prediction \hl{\cite{pei2020machine}}.

\subsubsection{Data augmentation} \label{Data_Aug}
Despite the rapid development in methodology and computing resources, DFT calculations are still computationally expensive and inefficient. First, the typical dataset size is limited to tens of thousands, beyond which the computing cost would be too high. The whole strategy of replacing DFT with an ML surrogate model would lose its merit. By comparison, in image recognition, the simple MNIST handwritten digits database already has 70,000 instances, while the popular ImageNet database for visual object recognition contains more than 14 million instances \cite{5206848, yang2019fairer}. Secondly, the atomic configuration space is enormous for multicomponent alloys \cite{DDJohnson_NatureCS}. For example, for a ternary system in a 1000-atom supercell, the total number of configurations is $3^{1000}$, which is much larger than the $3^{361}$ possible board configurations in the game played by the famous alphaGo \cite{Silver2017}.

Active learning is commonly employed to iteratively improve the dataset quality by incorporating important data points from simulations \cite{PODRYABINKIN2017171, PhysRevMaterials.3.023804}. In each iteration of active learning, a query algorithm is employed to determine the new data points, which are evaluated and added to the original dataset. There are different strategies for choosing the data points. For instance, one strategy is to recalculate the energies with DFT if the energies predicted by the MC simulations are outside the energy range of the original dataset \cite{LIU2021110135}. Similarly, in MD simulations, the trajectory-based iterative sampling algorithms \cite{doi:10.1063/1.466801, PhysRevMaterials.3.023804} can be used to sample the dynamically important regions in the configuration space and add new data points where the trajectories visit most frequently. Another strategy is to estimate the prediction error with multiple models and add new data when the prediction error is larger than a threshold value, which is demonstrated for neural networks \cite{C1CP21668F, PhysRevB.85.045439, TEICHERT2020113281} and Gaussian process \cite{PhysRevB.93.054112, doi:10.1063/1.5051772}. Finally, Shapeev {\it et al.} proposed an active learning strategy based on the so-called D-optimality criterion \cite{PODRYABINKIN2017171,PhysRevB.99.064114}, which selects the training set by maximizing the determinant of the information matrix. This method has been implemented in the MLIP package \cite{Novikov_2021}. 

Another method to expand the dataset is the generative adversarial network (GAN). GAN is a deep learning architecture comprised of generative and discriminative parts. In the training process, the generative model tries to produce faked data similar to the training data, while the discriminative model attempts to distinguish the true and faked data. A well-trained GAN should capture the training dataset's important latent variables, therefore serving as a powerful tool for data augmentation. For example, in a recent work, Lee et al. \cite{LEE2021109260} employed GAN to augment the dataset of HEA properties and demonstrate that such a technique can significantly improve the performance of a deep neural network model for phase prediction with the test accuracy enhanced from 84.75\% to 93.17\%. 

\subsection{Model selection} 
Commonly used descriptors for phase prediction include atomic number, atomic size, the entropy of mixing, enthalpy of mixing, electronegativity, valence electron concentration, elastic constants, melting temperature, and many others.
In general, the relevance of each feature to the labels is unknown beforehand, so a sufficiently large pool of features is needed to avoid the case that essential features are left out. However, this large set of features introduces redundancy and the curse of dimensionality, and a feature selection procedure is necessary to reduce the model complexity and the overfitting risk.

There are three categories of feature selection techniques. The first one is to use measures such as correlation coefficients, mutual information, or similarity scores to evaluate the features' independence and their correlation to the target variable. The second method uses regularization techniques such as LASSO regularization to penalize non-zero coefficients and, therefore, automatically construct a sparse model. Finally, a computationally more intensive method is to use different feature sets to train models in a brute-force manner and then use a hold-out set to measure the model performance and select the best set.

Other than the descriptors, the hyperparameters in a model can also be selected to achieve the best performance. The selection of model hyperparameters can be divided into two major categories. The first one is based on information criteria, such as the Bayesian information criteria (BIC) and Akaike information criteria (AIC) \cite{burnham2004multimodel}. The second one is to use hold-out samples (e.g., cross-validation) to measure the model performance and tune the hyperparameters. For such a purpose, grid search and random search are the two commonly used simple methods. The hyperparameters can also be tuned with more involved techniques, including  genetic algorithm \cite{ZHANG2020528,NatureCommCCAHard}, evolution strategies \cite{wierstra2014natural} \hl{\cite{ zhang2021enabling, zhang2021directional}}, gradient-based optimization, and Bayesian optimization methods \cite{snoek2012practical} such as Gaussian process and tree-structured Parzen estimator (TPE) \cite{bergstra2011algorithms}. 

\subsection{Model training}
Once an ML algorithm is selected, the ML model is completely determined by the optimal model parameters. The model parameters can be determined via the minimization of the loss function. An important point to note is that while the ML atomic interaction models are built to predict the atomic energy, in the training process, the target variable is the total energy computed by DFT methods. For an effective Hamiltonian, the total energy is the only target variable, while for interatomic potential, the atomic forces also need to be considered. Representing the total energies and atomic forces of the $k$-th configuration as $E^k$ and $F_i^k$ respectively, we calculate the mean squared error (MSE) loss function by
\begin{align}
{\rm{MSE}} = \frac{1}{N_{\rm{config}}}\sum_{k=1}^{ N_{\rm{config}}}\left[ \left( E^k_{\mathrm{ML}} -  E^k_{\mathrm{DFT}} \right)^2 + w_F^k \sum_{i=1}^{N_k}  \left( F^k_{i \mathrm{ML}} -  E^k_{i \mathrm{DFT}} \right)^2  \right],
\end{align}
where $N_{\rm{config}}$ is the number of configurations in the dataset, ${N_k}$ is the number of atoms in the supercell of the $k$-th configuration, and $w_F$ is the weight factor of atomic forces, which can be adjusted, e.g., according to the number of atoms in the supercell. Based on the ML model, the optimization of the loss function can be achieved with different algorithms, such as the ordinary linear regression, Newton's method, stochastic gradient descent (SGD), and Adam optimizer \cite{kingma2014adam}.

ML models are usually data-hungry. One interesting question is whether some form of local energies can be calculated by DFT and used for the fitting process instead of the total energies \hl{\cite{pasini2020fast}}. One obvious benefit of this method is that the number of data points can be greatly expanded from $N_{\rm{config}}$ to approximately $ N_{\rm{config}} \times N^{k}$.  This is particularly attractive for HEAs, where a large supercell is needed to represent the random configurations. The real space DFT methods, such as LSMS, are indeed feasible to evaluate the local contributions by integrating the energy density over the local Voronoi polyhedron. However, one problem with this method is that the local energies of neighboring atoms are highly correlated. For example, suppose there is a charge transfer between two neighboring atoms. In that case, the energy increase of one atom leads to decreasing the energy of the other, while the total energy change is much smaller than each individual atomic energy. As a result, although the atomic energy can be determined with a small relative error, it still gives rise to a large relative error for the averaged total energy, which has a standard deviation smaller than the local atomic energies.

\subsection{Result analysis}
After the ML models are trained, the next step is to evaluate their performance. For such purposes, hold-out samples that do not appear in the training dataset are usually used. For an objective and accurate evaluation, the testing samples should be representative and uncorrelated to the training samples; otherwise, it may give rise to a spuriously high test score. For classification tasks, quantities such as recall, precision, area under the curve (AUC), or receiver operating characteristic (ROC) curve are commonly used to measure the model performance. For regression tasks, the root mean square error (RMSE) and coefficient of determination ($R^2$ score) are widely used.

While well-trained ML models can be very good at making predictions, in many situations, the reasons for their effectiveness are also of great interest. 
For example, ML can be used to gain physics insights from the solutions provided by the model \hl{\cite{pei2019machine,pei2020machine,yin2021neural,pei2021mechanisms}}. Model interpretation is relatively easier for some ML models than others. For example, in decision trees, the importance of features can be evaluated by inspecting their effects on reducing the Gini index. For less transparent models, techniques such as the permutation feature importance can be employed to identify the most important features \cite{10.5555/1953048.2078195}. The permutation feature importance is defined by the decrease of the model performance when the selected feature's values are randomly shuffled. This is more challenging for neural networks. For simple neural networks, the sensitivity of the model with respect to the input features can be approximated by using the Taylor expansion to convert the nonlinear model into a linear one \cite{ML_Phase_sensitivity}. 

Deep neural networks are generally difficult to interpret due to their nonlinearity and high dimensions. One important technique to visualize high-dimensional data is the t-distributed stochastic neighbor embedding (t-SNE) algorithm \cite{van2008visualizing}. The t-SNE algorithm maps high-dimensional data into two or three-dimensional space where one data point keeps the similar data as its neighbors as in the original space. For classification, the latent variables obtained by t-SNE are visualized to infer the model's performance. If data with different labels are well separated in the reduced space, then the model can differentiate them easily; otherwise, the model is difficult to distinguish them \cite{LEE2021109260}.

\section{Atomic interaction models}
One important requirement for atomic interaction models is that they should be independent of the system size; thus, the model trained from small systems can be employed to investigate much larger ones. For such purpose, the  \textit{atomic energy} $E_i$ is a crucial quantity, which represents the energy contribution from an individual atom:
\begin{align}
E = \sum_{i=1}^{N_{\rm{atom}}} E_i({\boldsymbol \sigma}_i) + E_0, \label{eq:atomic_energy}
\end{align}
where ${\boldsymbol \sigma}_i=\{\sigma_i^0, \sigma_i^1, \sigma_i^2, \cdots \sigma_i^{N_n} \}$ represents the \textit{atomic environment} of the $i$-th atom, with $\sigma_i^0=(z_i, \mathbf{r}_i)$ represents the atomic species and position of the $i$-th atom, and $\sigma_i^{j\neq0} = (z_j, \mathbf{r}_{ij})$ represents the surrounding $N_n$ atoms within a cutoff radius $r_c$. $E_0$ represents the long-ranged electrostatic term that depends only on the chemical concentration, not atomic arrangement. As a result, the problem of establishing a mapping between configuration and total energy is reduced to mapping an atom's local environment to its atomic energy, as illustrated in Fig.~\ref{fig:AtomicEnv}. Note that when the atomic forces are also needed in the atomic interaction model, they should be fitted simultaneously with the local energies.

%\subsection{Descriptors}
%In this section, we first discuss the representation of chemical configurations, i.e., how to use symmetry-invariant quantities to encode the arrangement of atoms (local atomic environment), as illustrated in Fig.~\ref{fig:AtomicEnv}. There are two different approaches:
%\begin{itemize}
%\item {\it Symmetry-preserving (hand-crafted) descriptors}. This group of descriptors includes the Parrinello-Behler symmetry functions, power spectrum, bispectrum, moment tensor, etc.
 
%\item {\it End-to-end (automatically learned) descriptors}. Examples include the three-dimensional convolutional neural network (3DCNN) and graph convolutional neural network (GCNN).
%\end{itemize}
%We will have the following two sub-sections to offer the details.

\subsection{Symmetry-preserving descriptors} \label{symmetry_descriptor}
An encoding scheme that maintains the underlying symmetry needs to be devised to represent the atomic environment with high fidelity. As a categorical variable, the atomic species can be represented with one-hot encoding, element embedding \cite{ZhouE6411,yin2021neural}, or simply assigning a unique weight for each atomic species. A similar technique is used in constructing the atomic neighbor density function \cite{PhysRevB.87.184115}. The Cartesian coordinates can not be directly used as features since squeezing the positions into one-dimensional vectors would completely lose the underlying symmetry of the system. One simple scheme to address this problem is to use the radial distances $r_{ij}$ of atom $i$ to its neighboring atom $j$ as the input. An obvious drawback of this approach is that it ignores the angular distribution, which can be mitigated by adding the angle $\theta_{ijk}$ between the two lines formed by the relative positions $\mathbf{r}_{ij}$ and $\mathbf{r}_{ik}$. A well-known example is the Parrinello-Behler descriptor (also known as the symmetry function), which uses the product of a Gaussian function and a cutoff function to represent the radial arrangement of atoms. The function of $\cos(\theta_{ijk})$ is adopted to describe the angular distributions \cite{PhysRevLett.98.146401, https://doi.org/10.1002/qua.24890}. Other symmetry invariant quantities can also be used to represent chemical configurations, such as power spectrum and bispectrum \cite{PhysRevLett.104.136403, PhysRevB.87.184115}, and moment tensors \cite{doi:10.1137/15M1054183, Novikov_2021}. Another approach is to use physical observations to represent materials, such as the X-ray diffraction patterns \cite{YingZhangNC2020}.

\subsection{Machine-learning potentials}
Interatomic potentials are functions that describe the dependence of the potential energy on the atomic positions, which is crucial for molecular-dynamics simulations.
Popular traditional interatomic potentials include the Lennard-Jones potential, embedded-atom method (EAM) \cite{BECKER2013277}, and reactive force field (ReaxFF) \cite{ReaxFF}. Compared to these traditional potentials, ML potentials have two main features: Firstly, ML potentials take flexible rather than fixed function forms, and their accuracy can be systematically improved; Secondly, ML potentials adopt a data-driven approach comprised of training, validation, and tests, using dataset calculated by first-principles methods. As a result, high-quality ML potentials can approach the accuracy of quantum mechanical methods such as DFT and enable the description of systems that are too complex for traditional potentials depending explicitly on a few parameters. In the following, we will give a brief introduction to four commonly used ML potentials, i.e., the neural network potential (NNP) \cite{PhysRevLett.98.146401, https://doi.org/10.1002/qua.24890, Behler_2014, PhysRevLett.120.143001}, the Gaussian approximation potential (GAP) \cite{PhysRevLett.104.136403, PhysRevB.87.184115}, the spectral neighbor analysis potential (SNAP) \cite{THOMPSON2015316, doi:10.1063/1.5017641}, and the moment tensor potential (MTP)  \cite{doi:10.1137/15M1054183,  Novikov_2021}.

\subsubsection{High-dimensional neural network potential (HDNNP)}
Artificial neural networks (ANN) are widely employed to establish a nonlinear mapping between the atomic environment and the atomic energy. Typically a feed-forward multilayer perceptron (MLP) is used, comprising an input layer, a few hidden layers, and an output layer. A simple, fully connected layer is 
\begin{equation}
    \mathbf{Y} = \phi(\mathbf{W X} + \mathbf{b}), \label{eq:fully_connect_layer}
\end{equation}
where $\mathbf{X}$ is the input vector, $\mathbf{b}$ is the bias vector, $\mathbf{W}$ is the weight matrix, $\phi$ represents the activation function, and $\mathbf{Y}$ is the output vector. The vector $\mathbf{Y}$ is also the input for its next layer. In forward propagation, the atomic environment encoding mentioned is fed into the networks, and weight matrices in the hidden layers allow the network to learn atomic interactions at different hierarchical levels. The output layer predicts the atomic energies, which are summed over all atoms within the system to predict the total energy, as shown in Fig.~\ref{fig:AtomicEnv}. Since the neural networks are applied to every atom, NNP is also referred to as high-dimensional NNP (HDNNP) \cite{https://doi.org/10.1002/qua.24890}. The predicted energies are compared in the backward propagation with DFT results to calculate the loss function. By employing optimization algorithms such as the stochastic gradient descent (SGD), the parameters $\{\mathbf{W}, \mathbf{b} \}$ are then updated to reduce the expected loss. The forward and backward propagation is iterated until the model converges to satisfactory accuracy. Note that neural networks with different sets of $\{\mathbf{W}, \mathbf{b} \}$ parameters are used to represent the individual atomic species, but the atomic neural network for a given atomic species is the same. Different atomic energies are attributed to the different environments of atoms.

\subsubsection{Gaussian approximation potential (GAP)}
In GAP \cite{PhysRevLett.104.136403, PhysRevB.87.184115, Bartoke1701816, https://doi.org/10.1002/qua.24927}, the total energy is predicted by the Gaussian process regression (GPR) method that measures the ``similarity'' to the reference of atomic environments \cite{PhysRevX.8.041048}. Since the Gaussian process is computationally expensive for large datasets, a sparse representation \cite{10.5555/2976248.2976406} technique is employed to reduce the computational cost of the GPR from $O(N^3)$ to $O(N^2M)$, assuming the number of data points $N \gg M$. The choice of parameters for the kernel function $K$ reflects the assumed covariance between the random variables. They measure the similarity between atomic environments ${\boldsymbol \sigma}_i$ and ${\boldsymbol \sigma}_s$.
For example, if a dot-product kernel is used, the GPR is reduced to Bayesian linear regression. The atomic environment ${\boldsymbol \sigma}_i$ can be represented by the so-called \textit{atomic neighbor density}
\begin{align}
    \rho_i(\mathbf{r}) = \sum_{j} w_{Z_i}  f_{\rm{cut}}(r_{ij}) \exp(-|\mathbf{r}-\mathbf{r}_{ij}|^2/(2\sigma^2_{\rm{atom}})),
    \label{eq:neighbor_density}
\end{align} 
where $w_{Z_i}$ is a unique weight factor assigned according to the element of atom $i$, $\sigma_{\rm{atom}}$ is the smearing parameter, $f_{\rm{cut}}(r_{ij})$ is the cutoff function to ensure interactions beyond $r_c$ are ignored, and the summation of $j$ is over all atoms within the cutoff radius. Using Eq.~\ref{eq:neighbor_density}, the so-called smooth overlap of atomic positions (SOAP) kernel can be obtained by integrating the squared overlap of the neighbor densities of atom $i$ and $s$, over all possible 3D rotations \cite{PhysRevX.8.041048}. 
\begin{equation}
\tilde K({\boldsymbol \sigma}_i,{\boldsymbol \sigma}_s) = \int_{\hat R \in SO_3} d\hat R \left| \int d{\bf r} \rho_i({\bf r}) \rho_s (\hat R {\bf r}) \right|^2 \label{eq:kernel_matrix}
\end{equation}
In practice, the 3D rotation is accomplished analytically by expanding the atomic density in terms of spherical harmonics,
of which the integration can be expressed in terms of Clebsch-Gordan coefficients. The derivation details are well presented in Ref.~\cite{PhysRevX.8.041048} and are very similar to the SNAP, as discussed below. Once the matrix expression of the SOAP kernel is determined, the GAP can be obtained by solving the regression problem with the Gaussian process.

\subsubsection{Spectral neighbor analysis potential (SNAP)}
SNAP is closely related to GAP in that the central quantity is also the atomic neighbor density in Eq.~\ref{eq:neighbor_density}. For simplicity, it can be assumed that the density of the neighboring atoms for atom $i$ is made of a sum of atoms within the cutoff radius, as represented by the Dirac $\delta$ function,
\begin{align}
    \rho_i({\mathbf{r}}) = \sum_{j}^{r_{j}<r_c} \delta(\mathbf{r}_i-\mathbf{r}_{j}).  \label{eq:delta_density}
\end{align}
Expand the density using the spherical harmonics as the basis set, the above expression gives
\begin{equation}
\label{eq:density_expansion}
\rho_i({\bf r}) = \sum_{lm} c^i_{lm} Y_{lm}({\bf \hat r}),
\end{equation}
 with the expansion coefficients
 \begin{align}
 c^i_{lm} = \sum_{j}^{r_{j}<r_c} Y_{lm}({\bf \hat r}_{j}).
 \end{align}
The coefficients can be used to construct rotation-invariant descriptors since, in group representation theory, the direct product of two irreducible representations can be decomposed into a direct sum of irreducible representations.
The simplest one is the power spectrum
\begin{align}
{p}^i_{l} &= \sum^l_{m=-l} c^{i*}_{lm} c^i_{lm},
\end{align}
which is rotationally invariant since it commutes with the angular momentum operator. Similar to the power spectrum, more complex invariants such as bispectrum can be constructed using the radial distribution function rather than Dirac $\delta$ function and map the atomic neighbor density within $r_c$ to a sphere in four-dimensional space \cite{THOMPSON2015316}. The power spectrum and bispectrum in four-dimensional sphere can then be used as descriptors for various regression models, including the linear SNAP \cite{THOMPSON2015316}, quadratic SNAP (qSNAP) \cite{doi:10.1063/1.5017641}, and spectral neural network potential \cite{doi:10.1063/5.0013208}.

\subsubsection{Moment tensor potential (MTP)}
MTP uses the contraction of moment tensors to construct invariant descriptors. In MTP, the total energy is expressed as 
\begin{align}
    E = E_0 + \sum_{i=1}^{N_{\rm{atom}}} \sum_{\alpha} \xi_\alpha B_\alpha({\boldsymbol \sigma}_i),
\end{align}
where $\xi_\alpha$ are the coefficients that need to be determined from the dataset. The basis functions $B_\alpha$ are defined in terms of the moment tensor descriptors:
\begin{align}
    M_{\mu, \nu}({\boldsymbol \sigma}_i) = \sum_j f_\mu(r_{ij}, z_i, z_j) \underbrace{\mathbf{r}_{ij} \otimes \ldots \otimes \mathbf{r}_{ij}}_\text{$\nu$ times}.
\end{align}
The radial function $f_{\mu}(r_{ij},z_i,z_j)$ plays the role of ``selecting'' shells of atoms \cite{GUBAEV2019148}, and can be expanded in terms of a set of basis functions as 
\begin{equation}\label{eq:f_mu}
f_{\mu}(r_{ij},z_i,z_j) = \sum_{\beta=1}^{N_Q} c^{(\beta)}_{\mu, z_i, z_j}  f_{\rm cut}(r_{ij}) \varphi^{(\beta)} (r_{ij}),
\end{equation}
which is made up of the cutoff function $ f_{\rm cut}(r_{ij}) $ and the Chebyshev polynomials $\varphi_{\mu} (r_{ij})$. %within $r_{\rm min}$ and $r_c$. 
The tensor part $\mathbf{r}_{ij} \otimes \ldots \otimes \mathbf{r}_{ij}$ is comprised of the outer product of relative position vectors, which contains angular information of the atomic environment. $M_{\mu, \nu}$ can be interpreted as the $\nu$-th moments of inertia weighted by the $\mu$-th radial functions, and $c_{\mu, z_i, z_j}$ are parameters to be fitted from the data. The basis functions $B_\alpha$ are obtained from the contraction of moments $M_{\mu, \nu}$. The so-called level of moments can be defined in terms of $\mu$ and $\nu$ \cite{GUBAEV2019148}, then up to the level cutoff, all tensor moments are included in the local energy model. The MTP parameters $\theta=\{\xi, c\}$ are determined by minimizing the loss function for the training dataset. 

\subsection{Effective rigid-lattice Hamiltonian}
Compared to molecular dynamics, atoms in a rigid-lattice model are assumed to be located on lattice sites. During Monte-Carlo simulations, only the atomic species are updated according to a policy satisfying detailed balance, while the atomic positions are not relaxed. In a lattice model, the total energy can be expanded in terms of the cluster interactions
\begin{align}
E = E_0 + \sum_{\alpha} V_\alpha\phi_\alpha(\bm{\Sigma}) , \label{eq:cluster_energy}
\end{align}
where $\bm{\Sigma}$ represents one of the many possible configurations of a system, $\phi_\alpha$ represents the correlation function for the cluster $\alpha$ \cite{SANCHEZ1984334,pei2020statistics}. The correlation function is determined by (i) the different types of point, pair, triplet clusters, etc, (ii) the dimension of the cluster, and (iii) the atomic species. An excellent property of the correlation functions is they are orthogonal to each other in terms of the lattice sites and atomic species. $V_\alpha$ are the coefficients of cluster $\alpha$ known as effective cluster interactions (ECIs). The ECIs can be directly calculated with coherent-potential approximation (CPA) methods such as the generalized perturbation method (GPM) or the embedded cluster method (ECM), but the most popular approach is cluster expansion \cite{PhysRev.81.988, SANCHEZ1984334, NPJ_Widom}, which calculates the ECIs by fitting the DFT energies of selected configurations. Another effective atomic interaction model is the low-rank potential (LRP) \cite{SHAPEEV201726} proposed by Shapeev, which uses low-rank tensors in the MTP method as the expansion basis.

\begin{figure} [h]
    \centering
    \includegraphics[width=0.7 \linewidth]{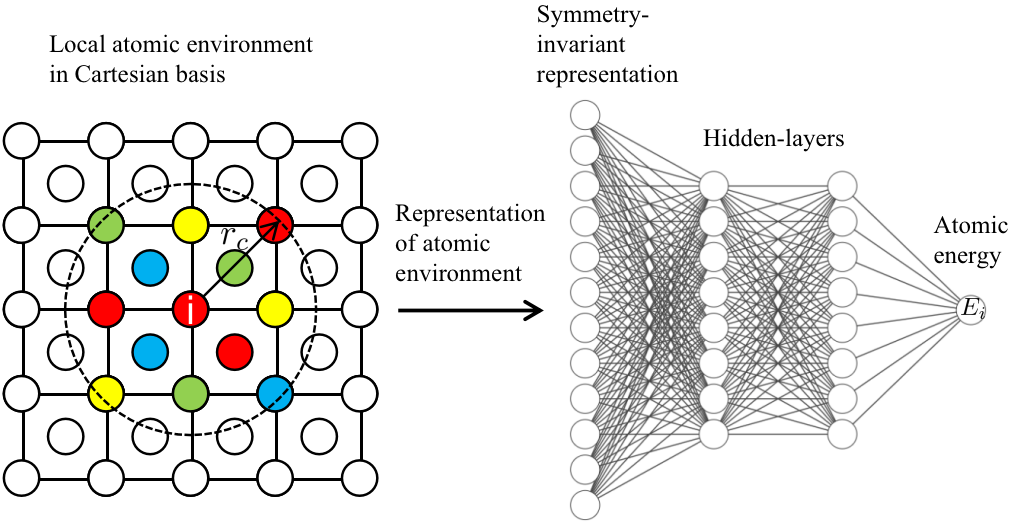}
    \caption{A schematic of a machine-learning model for energy calculations. (left) The atomic energy $E_i$ depends on the local atomic environment within a cutoff radius $r_c$. The different colors of the atoms represent different atomic species. The Cartesian positions and atomic species are converted into symmetry-invariant representations as inputs to ML atomic model. (right) A neural network is used as an example.}
    \label{fig:AtomicEnv}
\end{figure}

Despite its huge success in the theoretical study of alloys, the conventional cluster expansion has been challenged by HEAs due to a large number of interactions and the computational cost to extract the interaction parameters accurately \cite{PhysRevLett.116.105501, seko2009cluster, widom_2018}. A large number of interaction parameters is prone to overfitting, rendering the conventional structure-inversion method to determine ECIs risky. As a result, it is advantageous to adopt a data-driven approach for a robust model (Fig. \ref{fig:AtomicEnv}), assisted by techniques such as adding regularization terms and selecting suitable models. One simple example is to find the ECIs via regularized linear regression:
\begin{align}
\hat{\mathbf{V}} = \mathop{\arg\min}\limits_{\mathbf{V}} \;||\mathbf{E} -\boldsymbol{\phi V}  ||^2_2 + g(\boldsymbol{V}),
\end{align}
where $g(\boldsymbol{V})$ is the regularization term. One commonly used scheme for it is the ridge regularization, i.e.,
\begin{align}
g(\boldsymbol{V}) = \lambda || \boldsymbol{V} ||^2_2,
\end{align}
where the $\ell_p$-norm is defined as $|| x ||_p = (\sum_i |z_i|^p)^{1/p}$, and $\lambda$ is a regularization parameter that can be chosen to set the strength of penalizing large ECI values. Instead of choosing the $\lambda$ by hand, Bayesian ridge regression can be employed to automatically optimize $\lambda$ along with other parameters by maximizing the log marginal likelihood. Sometimes sparse solutions of the least square problem are preferred, where the $\ell_1$-norm LASSO (least absolute shrinkage and selection operator) regularization can be employed
\begin{align}
g(\boldsymbol{V}) = \lambda || \boldsymbol{V} ||_1.
\end{align}
The so-called group LASSO regularization is also introduced in Ref.~ \cite{PhysRevB.100.134108} to handle multicomponent systems. On the other hand, model selection can be employed to reach an optimal balance between accuracy and robustness. For example, Zhang {\it et al.} \hl{\cite{ZHANG2020108247}} proposed to use the Bayesian information criterion (BIC) for feature selection of the HEA energy model, 
\begin{equation}
\textup{BIC}_{\textup{RSS}} =n_d\log\left(\frac{\textup{RSS}}{n_d} \right)+k \log(n_d), \label{eq:BIC4}
\end{equation}
where $\textup{RSS}$ is the residual sum of squares, $n_d$ is the number of observed data, and $k$ is the number of parameters in the model. The first term represents the training error, the second term represents the model complexity, and a model is selected when $\textup{BIC}_{\textup{RSS}}$ reaches minima. Another extensively used model selection technique is the cross-validation score, which uses hold-out validation data to estimate the performance of different models. It is particularly useful when the dataset size is large. 

Another difficulty is that the linear assumption in Eq.~\ref{eq:cluster_energy} may not hold when only low-order terms are kept, and higher-order interactions are discarded, which is common for HEAs due to the $N^m$ scaling behavior of $N$-site clusters in an $m$-component system. To address this problem, nonlinear models such as kernel ridge regression, quadratic regression, and neural networks can be employed to model effective atomic interactions. For example,  Natarajan and Van der Ven \cite{ML_Cluster} developed a formalism that uses nonlinear models to calculate atomic energy. The local environment is represented with the so-called local correlation functions $G_\alpha^i$, and the total energy can be written as \cite{ML_Cluster}
\begin{align}
E = E_0 + \sum_{i} E_i(G_\alpha^i, G_\beta^i, \cdots), \label{eq:nonlinear_energy}
\end{align}
To describe the nonlinear effects, a fully connected 3-layer neural network consisting of 4, 4, and 2 nodes are used, respectively. The neural network model is trained with 1000 randomly generated distinct configurations and tested with 1346 distinct configurations. The model consistently outperforms the linear model, especially when the number of features is small. For example, for six local features, the test error of the NN model is 0.006 eV/atom, while the linear regression model gives an error of 0.01 eV/atom. 

\section{Machine-learning accelerated atomistic simulations}
\begin{figure} [h]
    \centering
    \includegraphics[width=0.8 \linewidth]{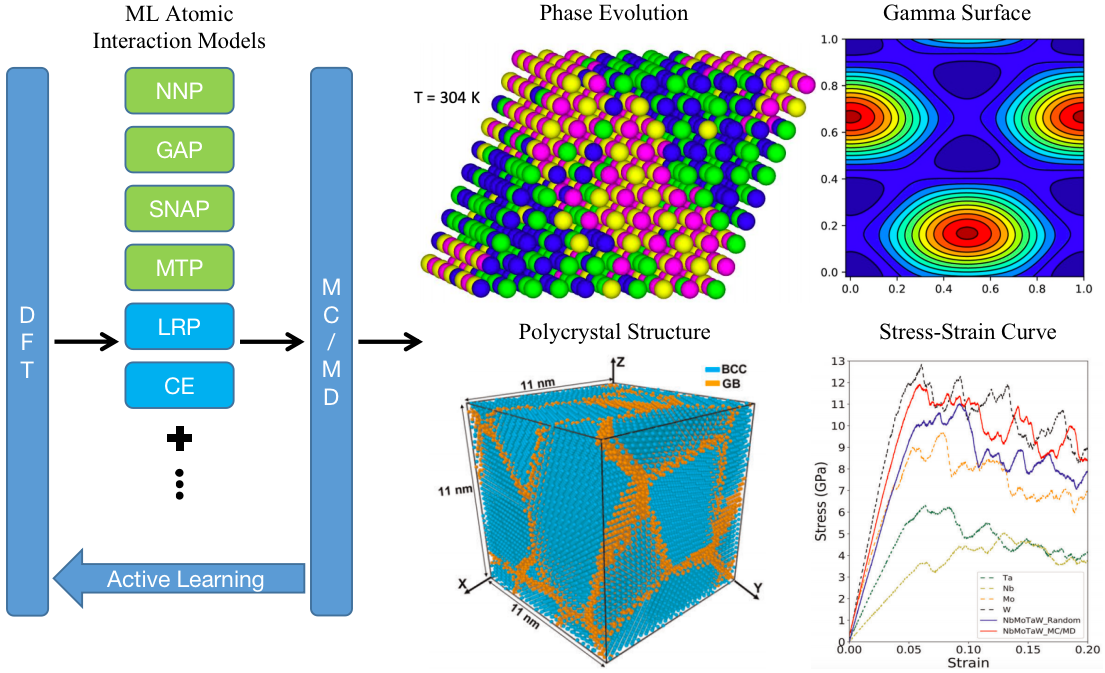}
    \caption{Schematic of atomistic simulations with ML models and their applications. (left) The scheme to train various ML models through active learning. (middle and right) Various thermodynamic and mechanical properties can be obtained from the simulations. The figures are reproduced from \cite{LIU2021110135, LIU2019107955, Ong_NPJ_SFE}.}
    \label{fig:Workflow_Atom}
\end{figure}

This section summarizes the progress of using ML surrogate models to accelerate the DFT-based atomistic simulations. These simulations can be divided into three categories: Monte-Carlo simulations, molecular dynamics, and hybrid MC/MD simulations. The main advantage of Monte-Carlo simulations is that, in principle, they can cover the entire phase space. This capability is crucial for studying CSRO and phase evolution at different temperatures. They are interesting not only because they can generate directly applicable results but also provide essential inputs for studying the temperature dependence of mechanical behaviors \cite{doi:10.1063/5.0064420}. One main limitation of Monte-Carlo methods is that they cannot describe non-equilibrium kinetic processes, such as the atomic motions under stress. On the other hand, molecular dynamics is good at studying atomic motions but typically only simulates a small portion of the phase space due to the limited timescale.
Hybrid MC/MD combines both methods by introducing MC swap steps to the MD simulations to increase the phase space visited. This acceleration is advantageous in many applications, although either MC or MD inevitably introduces some drawbacks to this hybrid method. If not suitably addressed, these drawbacks can also lead to erroneous results.
This section will also discuss ML-potential-based simulations for HEAs, as illustrated in Fig.~\ref{fig:Workflow_Atom}. The important applications are summarized in Tab.~\ref{tab:t2}.

\begin{table}[!ht] 
%\footnotesize
\centering
\caption{A summary of machine-learning applications in the atomistic simulations for multicomponent alloys. A few examples are given for pure metals to illustrate the methods. }
\label{tab:t2}
\begin{tabular} {p{0.2\textwidth}p{0.18\textwidth}p{0.3\textwidth}p{0.1\textwidth}} %{cccc }
\hline
\hline
%\hline
%\hline 
Materials & Methods & Physical Quantities & Refs \\ 
\hline
\hline
MoNbTaW & LRP+MC & Phase evolution& \cite{KormanMoNbTaW}   \\ \hline
 VCoNi & LRP+MC & Phase evolution & \cite{PhysRevMaterials.4.113802}  \\ \hline
AlNbTiV & LRP+MC & Phase evolution  & \cite{PhysRevMaterials.5.053803}  \\ \hline
 NiCoFeCr & LRP+MC & Phase evolution & \cite{MESHKOV2019106542} \\ \hline
 MoNbTaW, MoNbTaWV, MoNbTaWTi & Bayesian CE+MC & Phase evolution & \hl{\cite{ZHANG2020108247, LIU2021110135}}  \\ \hline
MoNbTaWV & GAP+ MC/MD & Defects, segregation &\cite{PhysRevB.104.104101} \\ \hline
CoFeNi & MTP+MD & Local lattice distortion, CSRO  & \cite{JAFARYZADEH20191054}  \\ \hline
TiZrHfTa & MTP+MD & Phase evolution, elastic constants & \cite{PhysRevMaterials.5.073801} \\ \hline
VZrNbHfTa & HDNNP+MD & Melt structure, radial distribution function & \cite{Balyakin_2020}  \\ \hline
MoNbTaW & SNAP+ MC/MD & Dislocation, GSFE, Peierls stress, stress-strain curve & \cite{Ong_NPJ_SFE}  \\ \hline
MoNbTaW & MTP+ MC/MD & CSRO on the mobility of dislocations & \cite{Asta_NC_MTP_2021} \\ \hline
MoNbTa & MTP+MD & Unstable SFE & \cite{hodapp2021machinelearning}  \\ \hline
Al-Cu & HDNNP+MD & GSFE & \cite{PhysRevMaterials.4.103601}  \\ \hline
Fe & GAP+ MD & Dislocation, GSFE, Peierls stress & \cite{PhysRevMaterials.2.013808, GAP_Fe_NPJ}  \\ \hline
Mo, Ta, Nb, W & SNAP+ MD & GSFE, Peierls stress & \cite{WANG2021110364}  \\ \hline\hline

\end{tabular}
\\ \raggedright
Acronyms in the table: low-rank potential (LRP), Bayesian cluster expansion (CE), Gaussian approximation potential (GAP), moment tensor potential (MTP), high-dimensional neural network potential (HDNNP), and spectral neighbor analysis (SNAP); chemical short-range order (CSRO), stacking fault energies (SFE), generalized stacking fault energies (GSFE); Monte Carlo (MC), molecular dynamics (MD), and hybrid MC/MD.
\end{table}

% %Examples: 
% \begin{itemize}
% \item ML models (e.g., low rank potential, Bayesian cluster expansion, GAP) plus Monte-Carlo simulations to investigate the order-disorder transitions in refractory HEAs \cite{KormanMoNbTaW, ZHANG2020108247, LIU2021110135, PhysRevMaterials.5.053803,  byggmastar2021modeling}, as shown in Fig.~\ref{fig:Workflow_Atom}.
% \item ML potentials (e.g., MTP, HDNNP) plus molecular dynamics to study local-lattice distortions, structural-phase changes, and partial-radial distribution of atoms \cite{JAFARYZADEH20191054,PhysRevMaterials.5.073801,Balyakin_2020}.
% \item ML potentials plus hybrid MC/MD to directly calculate the generalized stacking fault energies (GSFEs), dislocation cores, Peierls stresses, and stress-strain curves of HEAs, using large supercells unreachable with conventional DFT \cite{Ong_NPJ_SFE, Asta_NC_MTP_2021}, as shown in Fig.~\ref{fig:Workflow_Atom}.
% \end{itemize}

%We will address these simulations in the following sections.

\subsection{Monte-Carlo simulations}
Most of the fascinating mechanical properties of HEAs can be traced to the complex interplay of chemical order and disorder. At first-order approximation, random HEAs can be seen as made up of perfect crystals, with each site occupied randomly by one of the principal elements. In this simplified picture, random HEAs bear many similarities to pure metals. The first correction to this simple approximation is the local lattice distortion, induced by each atom's different chemical environments. Another important correction is the short-range and long-range order in the sample. As HEAs cool down from the melting temperature, atomic pairs with favorable bonding energy tend to appear more frequently, which gives rise to various degrees of order. These inhomogeneities impede the motion of dislocations, which can have a profound impact on strength and ductility. Therefore investigating the occurrence of order-disorder transition can provide essential insights into the exceptional mechanical properties of HEAs. 

\begin{figure} [h]
    \centering
    \includegraphics[width=0.8 \linewidth]{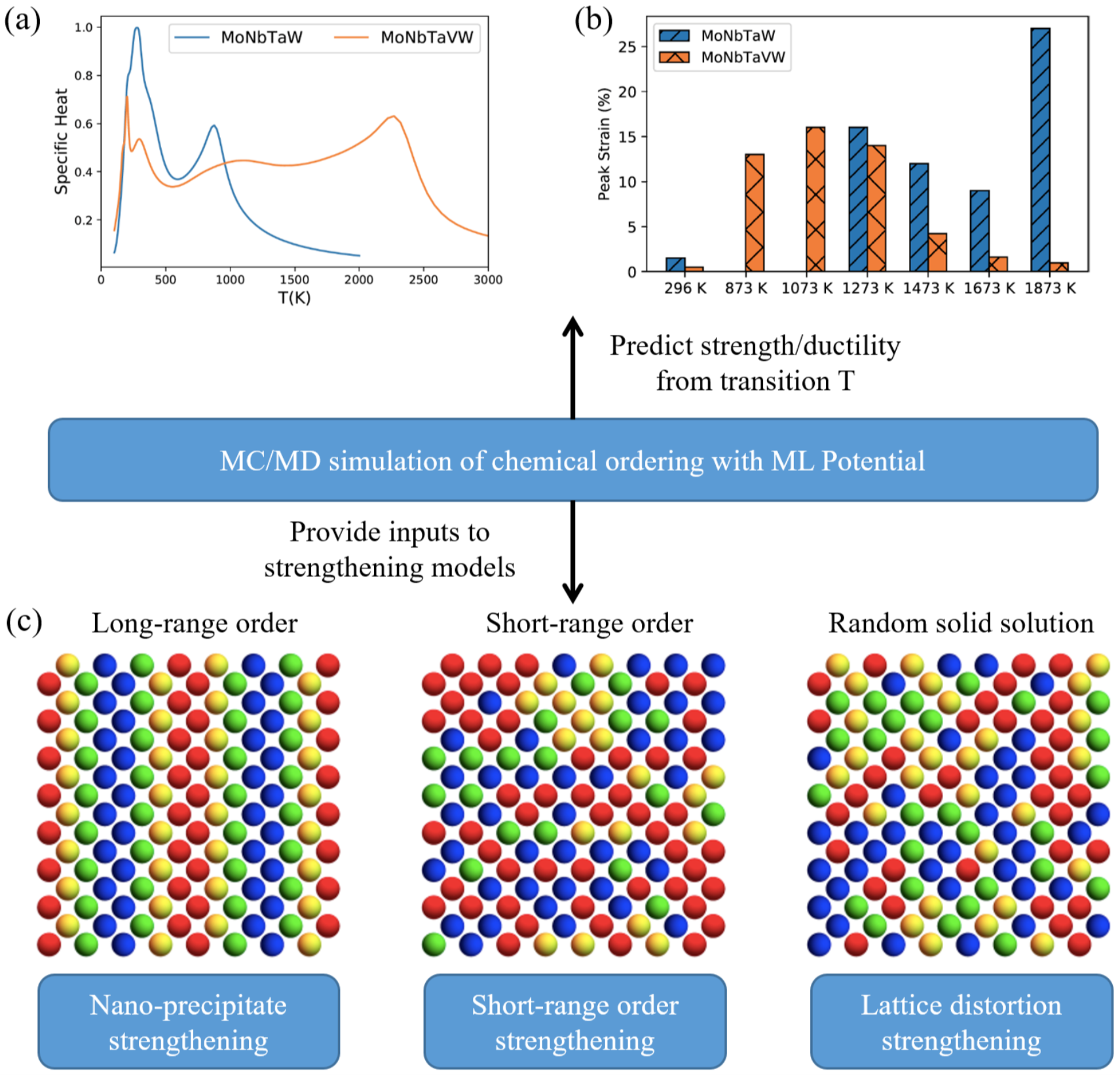}
    \caption{The chemical ordering in HEAs calculated by a machine-learning potential. Its relationship to strength and ductility is also explored. (a) Identification of order-disorder transitions in MoNbTaW and MoNbTaVW. (b) Experimental measurement of peak strains vs. temperatures in MoNbTaW and MoNbTaVW. (c) Illustration of the connections between different ordering and strengthening mechanisms. Figs. (a) and (b) are reproduced from Ref~\cite{LIU2021110135}.}
    \label{fig:SRO_HEA}
\end{figure}

Driven by such motivation, Liu {\it et al.} \cite{LIU2021110135} investigated three refractory HEAs: MoNbTaW, MoNbTaVW, and MoNbTaTiW using canonical MC simulations. The MC simulations were informed by an effective pair interactions (EPI) model. A data-driven approach was adopted for model selection, with the Bayesian information criterion \hl{\cite{ZHANG2020108247}} employed. They found the method produces highly accurate Hamiltonians that give error bars less than 0.1 mRy for all three HEAs. Moreover, the effective Hamiltonian accuracy was also validated by the total energies of 1000-atom MC configurations, which were calculated by the linear-scaling LSMS method. This demonstrates the importance of incorporating MC configurations in the training dataset through active learning. Furthermore, for each of the three HEAs, two order-disorder transition temperatures, i.e., $T_1$ and $T_2$ are identified from the computed specific heats and SRO parameters, as shown in Fig.~\ref{fig:SRO_HEA}. Finally, the calculations demonstrate that MoNbTaW has much lower order-disorder transition temperatures than MoNbTaVW, as illustrated in the specific heats vs. temperature plots in Fig.~\ref{fig:SRO_HEA} (a). Therefore, it is expected that in the range between 1000 K and 2000 K, MoNbTaW is primarily a solid solution, while MoNbTaTiW should contain a large fraction of second-phase precipitates. In other words, MoNbTaW is expected to be more ductile than MoNbTaVW in that temperature range, which is confirmed by the experimental results shown in Fig.~\ref{fig:SRO_HEA} (b).

There are also many other studies that combine machine learning models and Monte-Carlo simulations. In conjunction with rigid-lattice MC simulations, the low-rank interatomic potential (LRP) has been applied to compute the thermodynamics of several BCC and FCC multicomponent alloys. For example, Kostituchenko {\it et al.} investigated the order-disorder transitions in MoNbTaW \cite{KormanMoNbTaW}. The lattice distortion effect ignored in the rigid-lattice model was included by using DFT-relaxed structures. They found that the lattice relaxation is important for MoNbTaW when the temperature is below 500 K. Using the same approach, Kostituchenko and K\"ormann {\it et al.} also investigated the short-range order in BCC AlNbTiV \cite{PhysRevMaterials.5.053803} and FCC VCoNi \cite{PhysRevMaterials.4.113802}. For AlNbTaV, their calculations reveal a B2 ordering below 1700 K, which is mainly caused by the site preference of Al and Ti atoms. For VCoNi, they found an order-disorder transition around 1500 K, which was believed to be caused by the strong ordering tendency of Co-V and Ni-V pairs of the nearest neighbors. They also found that due to the relatively large size mismatch, neglecting the lattice relaxation can increase the predicted transition temperatures by more than 30\%. LRP has also been applied to study the FCC NiCoFeCr HEA \cite{MESHKOV2019106542} by Shapeev et al., where they found that Fe and Cr form sublattices at temperatures lower than 600$^{\circ}$C and 1230$^{\circ}$C, respectively.

\subsection{Molecular-dynamics simulations}
In addition to MC, molecular dynamics informed by ML models is also a common method for studying multicomponent alloys. Jafary-Zadeh {\it et al.} \cite{JAFARYZADEH20191054} investigated the local lattice distortion (LLD) effects in the CoFeNi base alloy through MD simulations with a moment tensor potential (MTP). The sample was melted at 2300 K for 10 ns, then quenched to 300 K, and kept in equilibrium for 5 nanoseconds. Based on the configurations obtained from MD simulations, they successfully separated the static, dynamic, thermal expansion, and chemical short-range order (CSRO) contributions to the LLD, as well as their impacts on the elastic properties. Gubaev {\it et al.} \cite{PhysRevMaterials.5.073801}  investigated the elastic constants of TiZrHfTa by chemically tuning the concentration of Ta, using MTP-informed MD simulations. They found that structural phase change can profoundly impact elastic properties. 
Other than MTP, high dimensional neural network potential (HDNNP) has also been applied to study the VZrNbHfTa melt \cite{Balyakin_2020}. The structure of this melt is calculated with both {\it ab initio} molecular dynamics (AIMD) and HDNNP. The results from AIMD and HDNNP were compared by analyzing the partial radial distribution functions, and a good agreement between the two methods was found. Furthermore, their simulation results show that vanadium atoms dislike other atomic species. Such an effect reduces the probability of forming a single-phase solid solution in this alloy. 

\subsection{Hybrid MC/MD simulations}

The generalized stacking fault energy (GSFE) is a critical quantity to describe plastic deformation behavior. It determines the core structure of dislocations and the minimal stress required for the onset of dislocation motions, i.e., Peierls stress. Extra energies are introduced when a perfect crystal is divided into halves by a slip plane and shifted along the slip direction of a slip system in question. The defected system is typically represented by a large supercell to avoid the spurious interaction between the faulted planes. A GSFE is calculated as the increased energy relative to the perfect crystal. Many different configurations need to be evaluated to account for the chemical disorder, further increasing the computational cost. Therefore, it is highly desirable to use efficient ML potentials to accelerate the GSFE calculations for multicomponent alloys.

\begin{figure} [h]
    \centering
    \includegraphics[width=1.0 \linewidth]{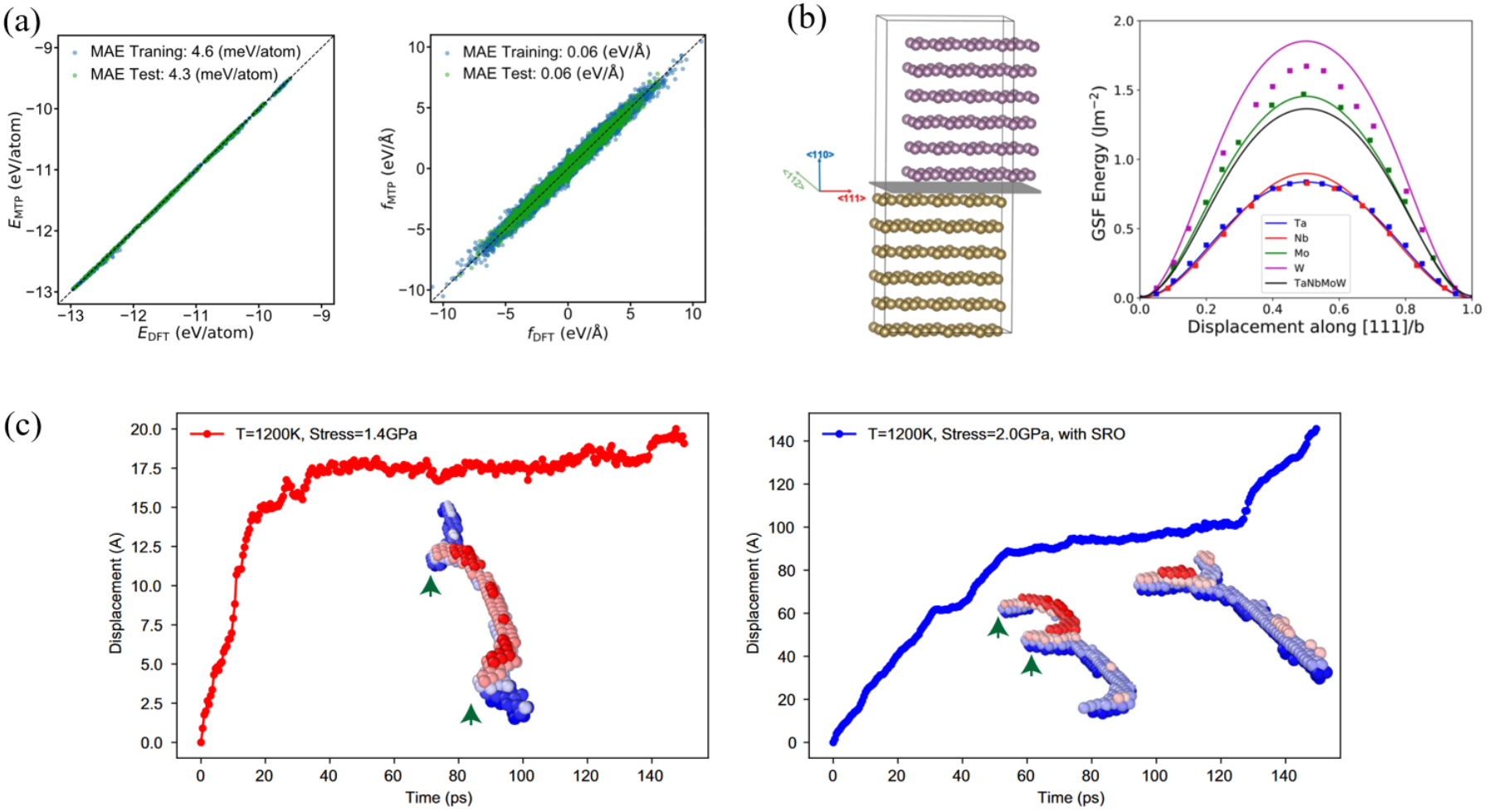}
    \caption{Applications of ML potentials to simulate dislocation properties. (a) Comparison of the energies and forces calculated from an ML model with those from DFT. (b) Comparison of the GSFEs calculated from a SNAP potential (lines) and DFT (square markers) for metals and NbMoTaW in special quasi-random structures. (c) Displacement vs. time curve of a screw dislocation under shear stress. The left subfigure corresponds to a random sample at 1.4 GPa and 1200 K, while the right one corresponds to an SRO sample at 2.0 GPa and 1200 K. Figs. (a) and (c) are reproduced from Ref.~\cite{Asta_NC_MTP_2021} and (b) from Ref.~\cite{Ong_NPJ_SFE}, under a Creative Commons CC BY license. }
    \label{fig:MoNbTaW_SFE}
\end{figure}

Ong {\it et al.} investigated the strengthening mechanisms in MoNbTaW with hybrid Monte-Carlo/molecular-dynamics (MC/MD) simulations informed by a SNAP model (Fig. \ref{fig:MoNbTaW_SFE}) \cite{Ong_NPJ_SFE}. The GSFEs are calculated with a large supercell containing 36,000 atoms, which is beyond the capability of conventional DFT. They found that the GSFEs of MoNbTaW are closer to those of W and Mo but much larger than those of Ta and Nb. They also determined the dislocation core structures by directly inserting a dislocation into the supercell and relaxing the structure with the SNAP model. The Peierls stress was then measured when the dislocation started to move. The Peierls stresses' values are 1620 $\pm$ 637 MPa for screw dislocations and 320 $\pm$ 113 MPa for edge dislocations. Finally, they also evaluated the uniaxial compressive stress-strain behavior of MoNbTaW polycrystals, with segregation to the grain boundary and SRO distributions obtained from MC/MD simulations. In addition, the MTP potential is also employed to study the impact of CSROs on dislocation mobility in the refractory HEA MoNbTaW, using a hybrid MC/MD simulation with a huge supercell of 573,672 atoms \cite{Asta_NC_MTP_2021}. 
%(\url{(https://creativecommons.org/licenses/by/4.0/})
% under a Creative Commons CC BY license (\url{(https://creativecommons.org/licenses/by/4.0/})

Using an MTP potential, Hodapp and Shapeev calculated the $1/4$[111] unstable stacking fault energies (SFEs) of Mo-Nb-Ta alloys with varying chemical concentrations \cite{hodapp2021machinelearning}. They demonstrated that the unstable SFE predicted by MTP is in excellent agreement with DFT, with a maximum relative error of $6$-$7\%$. They also showed that the DFT training data could be reduced from the order of 10,000 to about 50-100, assisted by active learning. This feature renders the MTP method attractive for the high-throughput screening of multicomponent alloys compared to the traditional DFT methods.

Curtin {\it et al.} employed HDNNP and kinetic MC simulations to study the dislocations and GSFEs of several alloys, including Al-Cu \cite{PhysRevMaterials.4.103601}, Mg alloy \cite{PhysRevMaterials.4.103602}, and Al-Mg-Si \cite{PhysRevMaterials.1.053604}. For Al-Cu, they calculated the GSFEs of the $\theta$ and $\theta''$ precipitates using HDNNP, a type of angular-dependent potential (ADP, an extension of EAM potential), and DFT. The results demonstrated that the neural network potential better agrees with the experimental results compared to the ADP EAM potential.

In a more recent study, Byggm\"astar {\it et al.} \cite{PhysRevB.104.104101}  developed a tabulated Gaussian approximation potential for MoNbTaWV and used it to study the radiation damage effects via hybrid MC/MD simulations. In the bulk alloy, they observed an ordering of Mo-Ta and V-W, consistent with other studies \cite{LIU2021110135}. They also found that vanadium in the damaged alloys segregates to the compressed interstitial-rich regions.

In addition to multicomponent alloys, ML potentials are also developed to study the GSFEs of pure metals, which is more straightforward for comparison with DFT methods. Dragoni {\it et al.} calculated the GSFEs in BCC iron with a GAP model trained with 150,000 local atomic environments, and the results are in excellent agreement with that of DFT  \cite{PhysRevMaterials.2.013808}. The GAP model outperforms EAM potentials. Maresca {\it et al.} further investigated dislocation cores in BCC iron with the GAP model \cite{GAP_Fe_NPJ}. The Peierls stress was calculated to be about 2 GPa, consistent with the DFT prediction. Using a SNAP potential, Wang {\it et al.} studied the GSFEs and Peierls stresses in refractory BCC metals Mo, Ta, Nb, and W  \cite{WANG2021110364}. Reasonable agreement with DFT was achieved for all the metals.

\begin{figure} [h]
    \centering
    \includegraphics[width=1.0 \linewidth]{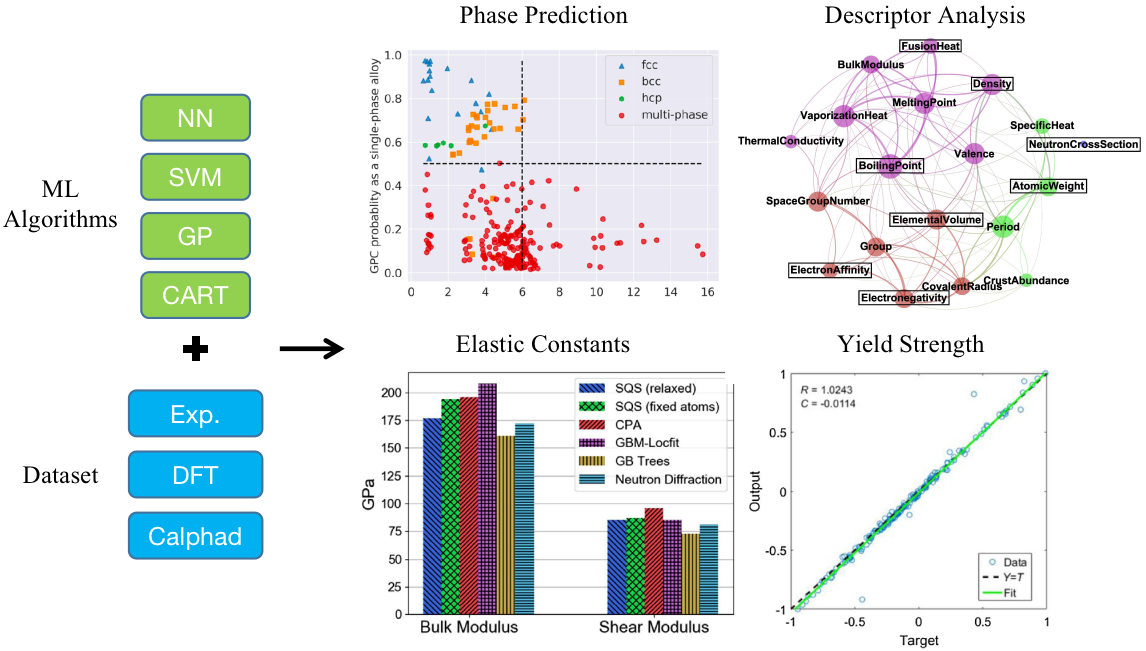}
    \caption{Machine-learning models for macroscopic properties. (left column) The essential components of machine-learning models; (middle and right) The thermodynamic and mechanical properties that can be calculated from the models. For a better understanding of the physics of the problem, relations between the descriptors are analyzed. The figures are reproduced from Ref.~ \hl{\cite{pei2020machine}} (phase prediction), \hl{\cite{pei2019machine}} (descriptor analysis), \cite{KIM2019124} (elastic constants), and \cite{ZHENG2021156} (yield strength).}
    \label{fig:Application_bulk}
\end{figure}

\section{Physical-property prediction}
In this section, we will discuss the applications of ML algorithms to predict physical properties. Important cases are summarized in Tab.~\ref{tab:thermo} and Tab.~\ref{tab:mechanical}. In short, we will discuss the following aspects in this section:

\begin{table}[!ht] 
%\footnotesize
\centering
\caption{A summary of machine-learning applications for the predictions of phases in HEAs.}
\label{tab:thermo}
\begin{tabular} {p{0.18\textwidth}p{0.18\textwidth}p{0.3\textwidth}p{0.05\textwidth} p{0.05\textwidth}}%{lllcccc }
\hline
\hline
%\hline
%\hline 
Methods & Target & Dataset \& Performance & Refs & Year \\ 
\hline
\hline
GP & SS formation & AUC = 0.97, 1252 data,  &\hl{\cite{pei2020machine}} & 2020   \\ \hline
GP & SS formation & >300 data & \cite{TANCRET2017486} & 2017 \\ \hline
KNN, SVM, NN & Phase & 74.3\% accuracy (NN), 401 data  & \cite{HUANG2019225} & 2019\\ \hline
NN & Phase  & over 80\% accuracy, 118 data & \cite{ISLAM2018230} & 2018\\ \hline
SVM & Phase & over 90\% accuracy, 322 data &\cite{PhysRevMaterials.3.095005}  & 2019 \\ \hline
DNN+GAN & Phase & 93.17\% accuracy, 989 data + GAN augmented data & \cite{LEE2021109260} & 2021 \\
\hline 
Model selection with GA & Phase & 88.7\% accuracy for solid solution formation and 91.3\% for crystal structure & \cite{ZHANG2020528} & 2020\\  \hline
Gradient-boosting & Phase and Young's modulus & 61\% accuracy for phase, 0.1\% to 42\% error for Young's modulus, 329 data & \cite{ROY2020152} &2020\\ \hline
Gradient-boosted DTs & Phase & 96.41\% for predicting single-phase solid solution, 1807 data & \cite{YAN2021110723} & 2022 \\ \hline 
Random forest & Phase &  100\% training accuracy, 134 data & \cite{KAUFMANN2020178} & 2020 \\ \hline
NN & Phase & Pearson's R=0.983, 321 data & \cite{WU2020278} & 2020 \\ \hline

\hline
\end{tabular}
\\ \raggedright
Acronyms in the table: Gaussian process (GP), K-nearest neighbors (KNN), neural network (NN), support vector machine (SVM), deep neural network (DNN), generative adversarial networks (GAN), genetic algorithm (GA), decision trees (DTs), the area under the curve (AUC); Solid solution (SS).
\end{table}

%Examples:
\begin{itemize}
\item Use ML algorithms to predict solid solution, intermetallics, amorphous, and mixed phases \cite{ZHANG2020528, HUANG2019225,ISLAM2018230, TANCRET2017486}\hl{ \cite{pei2020machine}}, as shown in Fig.~\ref{fig:Application_bulk}.

\item  Use ML algorithms to predict the crystal structure, such as FCC, BCC, HCP \cite{KAUFMANN2020178, PhysRevMaterials.3.095005,LEE2021109260}.

\item Identify important descriptors and design new prediction rules with the help of ML \cite{ML_Phase_sensitivity}\hl{\cite{ pei2019machine, pei2020machine}}.

\item Use ML algorithms to predict the mechanical properties of HEAs \cite{NatureCommCCAHard, WEN2019109, peng2020coupling}.
\end{itemize}

%We will have the following sub-sections to offer thorough details.

\subsection{Prediction of solid solutions and multi-phases}
Thermodynamics determines the phases in alloys, which provides essential information to design high-performance alloys, including HEAs. Initially, HEAs were defined as equiatomic single-phase multicomponent alloys; therefore, many pioneering HEA studies focused on phase formations. The published thermodynamic data renders the ML predictions for phase formation possible, which is an active research area at the boundary of ML and HEAs. Below we will present some representative examples and briefly summarize a number of other studies.

Usually, we need to predict the phases for a given alloy composition. Its inverse problem, i.e., identifying the alloy compositions that meet specific targets \hl{\cite{Pei2021AS}}, is even more interesting. The targets can be thermodynamic, mechanical, or other physical properties. The vast compositional space of HEAs is largely unexplored, and the synthesized HEAs are only a tiny fraction of the literally uncountable HEAs. This feature renders an efficient exploration of the HEA composition-phase space particularly interesting. Actually, there are some exciting proceedings in this direction \cite{ABUODEH201841}. The inverse problem for alloy design is mathematically equivalent to a constraint satisfaction problem that can be solved using more efficient ML techniques, e.g., supportive vector domain description. However, there is still a lack of reliable and efficient methods to inform the algorithms and validate their predictions. The CALculation of PHase Diagram (CALPHAD) method can be a good choice when experimental thermodynamic data become increasingly available to inform and validate the CALPHAD method. Another choice would be {\it ab initio} thermodynamics, given that new efficient algorithms are constantly proposed and powerful supercomputers have become increasingly more affordable.

High-quality (the most relevant) features and suitable ML algorithms are the preconditions for the success of ML models in phase prediction in HEAs. Many descriptors can be used as features for ML models. Zhang \textit{et al.} performed a comprehensive study that involved many descriptors and ML algorithms, making a total of 45 ML models \cite{ZHANG2020528}. Using genetic algorithms (GA), they selected the best ML model and material descriptors from a large model pool and demonstrated their efficiency on two-phase formation problems in HEAs. The optimized model demonstrates an accuracy of 88.7\% for identifying solid-solution and non-solid-solution phases and an accuracy of 91.3\% in predicting body-centered-cubic (BCC), face-centered-cubic (FCC), and dual-phase HEAs. Similarly, Huang \textit{et al.} also predicted the phase formations in HEAs \cite{HUANG2019225,ISLAM2018230}. They compared the accuracy of the K-nearest neighbors (KNN), support vector machine (SVM), and artificial neural network models on the prediction of the solid solution (SS), intermetallic (IM) compound, and mixed SS and IM phases. They used an experimental dataset of 401 HEAs, including 174 SS, 54 IM, and 173 SS+IM phases. The ANN model is the most accurate, with a value of 74.3\%, compared to 68.6\% and 64.3\% for KNN and SVM.

Identifying high-entropy solid solutions is one of the primary research interests among the HEA community. There are a large number of ML studies on this topic, which render it difficult to summarize all of them exhaustively. Here we can only briefly summarize several of them.
Based on a critical assessment and a Gaussian process statistical analysis, Tancret \textit{et al.} proposed a robust strategy to predict the formation of solid solutions \cite{TANCRET2017486}. They took into account most of the previously proposed criteria simultaneously. The method can be readily used as a guide to design new solid-solution HEAs. Lee \textit{et al.} predicted AM, SS, IM, and mixed IM and SS phases with a deep neural network \cite{LEE2021109260}. The hyper-parameters for the regularized deep neural network are searched via Bayesian optimization\cite{snoek2012practical}. To obtain the large dataset needed for neural network models, they employed conditional generative adversarial networks (GAN) to generate additional data (data augmentation), which can improve the performance of the neural network model. They also demonstrated that the optimized neural network model could reach an accuracy of 84.75\%, performing superior to SVM, decision tree, and XGBoost models.
Roy \textit{et al.} predicted Young's modulus, as well as phase identification, for a range of multicomponent alloys \cite{ROY2020152}. They use gradient-boosting decision trees to predict the phases and Young's modulus. The agreement is generally good for dozens of multicomponent alloys in the Mo-Ta-Ti-W-Zr family. Kaufmann \textit{et al.} also evaluated the performance of random forests for predicting solid-solution formations \cite{KAUFMANN2020178}. 

%\subsection{Prediction of the crystal structures}
A finer level of the problem is to predict the crystal structures of the formed solid solutions. Li \textit{et al.} \cite{PhysRevMaterials.3.095005} use support vector machine model to distinguish different crystal structures. The trained model is employed to make predictions of the crystal structure for a large compositional space made up of 16 elements as an effort to design new alloys. A summary of the above examples is given in Tab.~\ref{tab:thermo}.
%(support vector machine model, distinguishing stable BCC and FCC HEA phases). 

\subsection{Machine-learning informed new rules for phase predictions}

In addition to predicting phase formation, ML can also offer physical insights into the role of elements in phase formation. For example, in one study, researchers uncovered the eutectics design by ML in the Al-Co-Cr-Fe-Ni high-entropy system. ANN models classified the roles of elements \cite{WU2020278}. In the Al-Co-Cr-Fe-Ni system, a database of 321 alloys was built up using the literature results and CALPHAD calculations. It contains compositions and phase constitutions of these alloys. The ANN model was then trained to predict many near-eutectic compositions. The findings from the AI model can be used to formulate a new rule of thumb for designing high entropy alloys.
\begin{figure} %[h]
    \centering
    \includegraphics[width=0.9\linewidth]{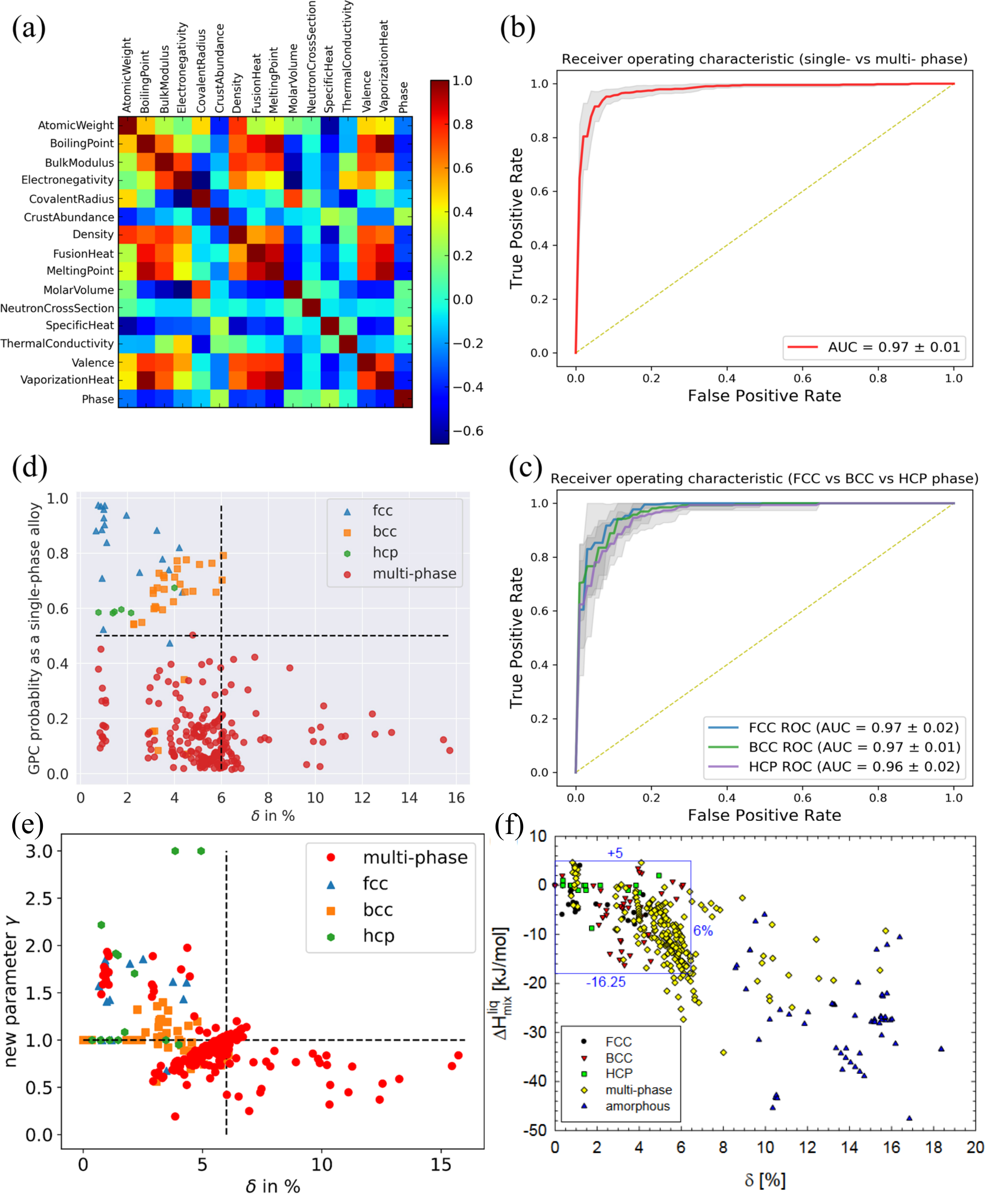}
    \caption{The machine-learning informed prediction of solid solutions. (a) The correlation matrix of elemental properties and alloy phase. The matrix shows that simple pair correlation is not enough to capture the mapping from elemental properties to alloy phases well. (b) The receiver operating characteristic (ROC) of Gaussian process classification with confidence band for single-phase versus multi-phase. (c) The receiver operating characteristic (ROC) of Gaussian process classification with confidence band for face-centered cubic (FCC) versus body-centered cubic (BCC) versus hexagonal closest packed (HCP) single-phase classification, respectively. The area under the curve (AUC) is calculated for each ROC curve. (d) Gaussian process classification (GPC) probability as a single-phase alloy versus atomic size difference. The symbols (triangle, square, hexagon, and circle) represent the experimentally measured phase states. (e) Prediction of the new rule $\gamma \ge 1$ the new rule alone gives an accuracy of 73\%, but with the lattice misfit rule, it slightly increases to 81\%. For better visualization, the values of $\gamma > 3$ are shifted to 3. (f) The same experimental data points cannot be well classified by previously proposed rules. Figures (a)-(e) are reproduced from Ref.\hl{\cite{pei2020machine}}, and (f) is reproduced from Ref.\cite{gao2017thermodynamics}.}
    \label{fig:npjCM20}
\end{figure}

We provide a detailed example to elaborate our discussion on the physical insights that ML can offer. Specifically, we demonstrate how to use ML to improve the prediction of thermodynamic properties of HEAs. Different from all above-mentioned examples, this example goes beyond the traditional ML framework, and demonstrates how ML can help construct a new thermodynamics-based rule to predict the formation of high-entropy solid solutions (Fig. \ref{fig:npjCM20}) \hl{\cite{pei2020machine}}. We adopt the Gaussian Process Classification (GPC) algorithm for this purpose. GPC algorithm is a Bayesian method that constructs a distribution based on existing data. Since it provides distribution for each prediction rather than a number, it can mitigate overfitting, a common problem for ML. When the data volume is large, it yields a distribution with a small variance; when the data volume is small, the variance is large. This feature renders it suitable for a small dataset size, which is usually the case for materials science problems. We collected 1252 multicomponent and binary alloys whose phases (BCC, FCC, HCP, or multiphase) are known from the literature. Many of these alloys have been described by Gao {\it et al.} \cite{gao2017thermodynamics}. Alloys with single phases are labeled as "1", whether BCC, FCC, or HCP. The multiphase alloys are labeled as "0". %These labels enter our GPC algorithm as the output. 
The features are constructed based on the 85 elemental properties available for relevant elements in the Periodic Table of Elements \cite{periodictable}. The physical properties of each multicomponent alloy are taken as the averaged elemental properties over their concentrations. %The properties that are obviously irrelevant to the phase formation are excluded from consideration. 
The properties are brutal-force correlated with the phase states, and some important ones are shown in Fig. \ref{fig:npjCM20}(a). The performance of our GPC model is measured by the so-called area under the curve (AUC) measure (see \ref{fig:npjCM20}b-c). The model's overall performance yields a very high AUC of 0.97; more detailed performance for each involved crystal structure is close, which is 0.96-0.97, showing the performance is not biased among the different crystal structures.

The model performance is more clearly seen by its accuracy of prediction. The criterion is that an alloy with a probability larger than 0.5 is considered to be a solid solution. Half of the 1252 alloys are randomly picked for model training and the others for validation with an accuracy of 93\%. The predictions are shown in Fig. \ref{fig:npjCM20}(d) with the GPC criterion as one of the axes. Almost all solid solutions, either FCC, BCC, or HCP, all have a probability larger than 0.5, in contrast to the multiphase alloys. For better visualization, the quantity of the commonly used lattice distortion criterion is added. Alloys with lattice distortion $\delta \le 6\%$ are generally considered to form solid solutions. 

The ML results significantly improve over the traditional criterion based on the enthalpy of formation $\Delta H$. The enthalpy of formation measures the thermodynamic tendency of the atoms to bind together. Too positive $\Delta H$ cannot bind atoms together (phase separation), while too negative $\Delta H$ will form intermetallics rather than a solid solution. The favorable energy window is $\Delta H \in [-16.25,5]$ kJ/mol for solid solutions. However, this empirical rule is inaccurate since many multiphase alloys also have formation energy in this energy window, as is shown in Fig. \ref{fig:npjCM20}(f). Compared to the empirical rule, the machine-learning model is indeed very encouraging.

As demonstrated by the ML model, the solid solutions can be accurately determined by the set of features that define a complex high-dimensional space. The inaccuracy of the one-dimensional empirical rule must originate from the inappropriate projection direction of the high-dimensional space. The next question would be, can we construct a more accurate rule informed by the ML results and the thermodynamic laws? We show this is indeed possible. We do not offer the details here but show the thermodynamics-based rule we derived in Ref. \hl{\cite{pei2020machine}}. With the most relevant physical properties identified by ML, we find solid solutions have a so-called $\gamma \ge 1$, with $\gamma$ defined as
\[
\gamma :=
  \begin{cases}
\Delta G_{N}/\min(\Delta G_{2}) & \text{if~} \min(\Delta G_{2})<0; \\
-\Delta G_{N}/\min(\Delta G_{2}) & \text{if~} \Delta G_{N}<0 \text{~and~} \min(\Delta G_{2})>0. 
  \end{cases}
\]
$\Delta G$ is the free energy of multicomponent and binary alloys. They can be analytically calculated, and their definition is referred to Ref. \hl{\cite{pei2020machine}}. The rule $\gamma \ge 1$ together with the lattice distortion $\delta$ parameter is shown in Fig. \ref{fig:npjCM20} (e). The overall accuracy of 75\%. The machine-learning informed new rule is nonetheless less accurate but physically more transparent and convenient for high-throughput screening in new alloy design.

\subsection{Machine-learning based order parameters}
Order parameters are used to describe the phase states of materials. It is one of the central concepts of condensed matter physics and materials science \cite{nix1938order,Cowley1950,OWEN2016155}. ML can provide a comprehensive scalar order parameter that has its unique advantages \cite{yin2021neural}. This can also be deemed as one of the contributions of ML to materials physics.

\begin{figure} %[h]
    \centering
    \includegraphics[width=0.95 \linewidth]{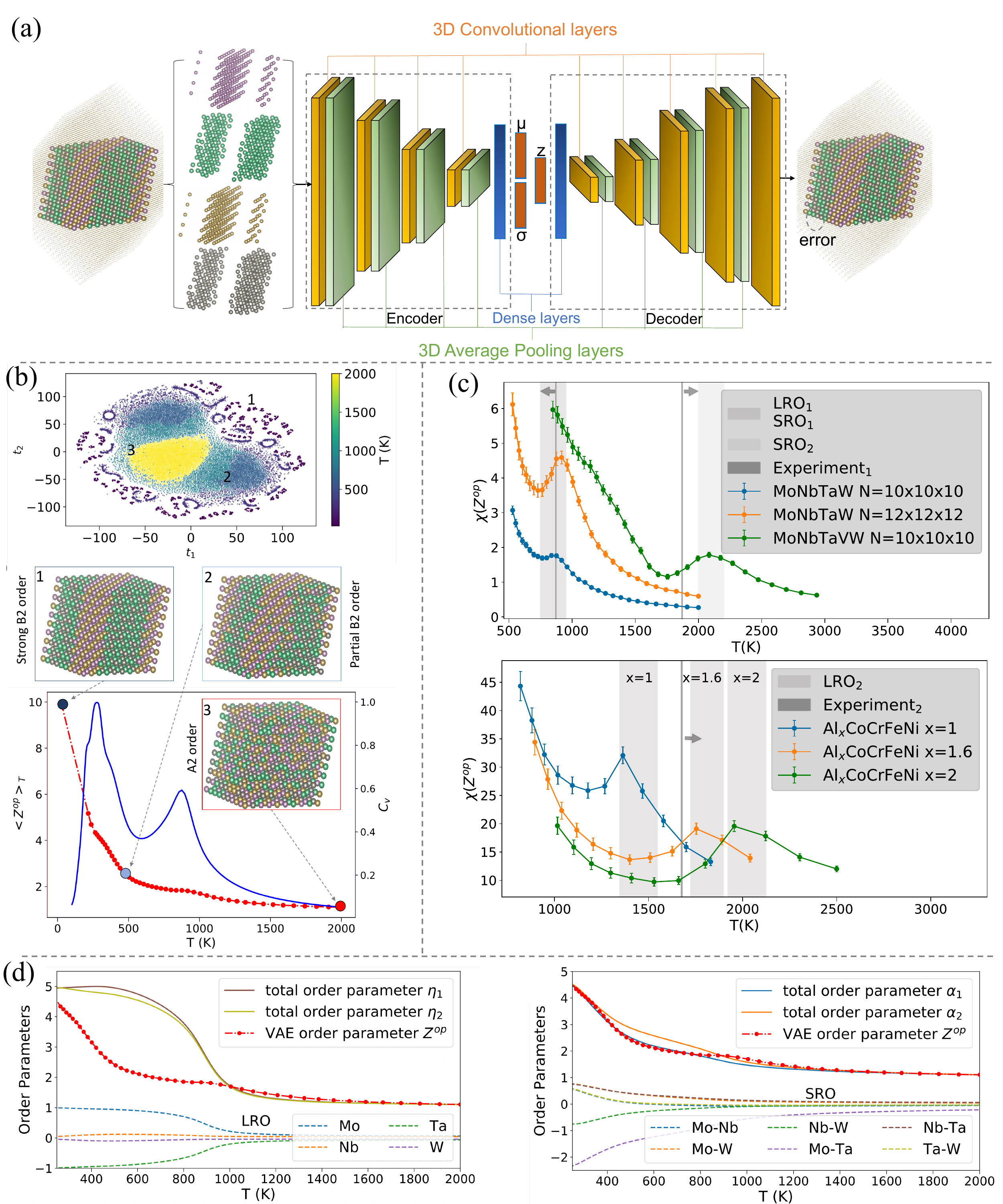}
    \caption{Neural network-based order parameter. (a) The neural network architecture of the VAE model. Both the encoder and decoder consist of several combinations of a 3D convolutional layer and a pooling layer, except the last convolutional layer for output. The latent variable $z$ is sampled from a Gaussian distribution with mean $\mu$ and standard deviation $\sigma$. (b) The 2D t-SNE plot of VAE 12D embedding on the test data. $t_1$ and $t_2$ are t-SNE embedding variables transformed from the 12D latent variable $z$. The inserted snapshots are sampled configurations at three different temperatures. (c) Comparison of the VAE order parameter with experimental data. The experimental data validate the performance of the VAE order parameter. (d) Comparison between the VAE order parameter and conventional order parameters. (left) The LRO for each element and the total order parameters in terms of LRO. (right) The SRO for each element pair and total order parameters in terms of SRO. This figure is reproduced from Ref. \hl{\cite{yin2021neural}}.}
    \label{fig:HEA-order-parameter}
\end{figure}

%We proposed a neural network-based order parameter. 
We elaborate on this idea by taking one of our recent studies as an example (see Fig. \ref{fig:HEA-order-parameter}). The neural network architecture is constructed based on the variational autoencoder (VAE) and shown in Fig. \ref{fig:HEA-order-parameter}(a). Both the encoder and decoder consist of several composite layers, except the last convolutional layer next to the output. Each composite layer consists of a 3D convolutional layer followed by an average pooling layer. Embedded in two dense layers is the latent variable $z$ that is sampled from a Gaussian distribution with a mean $\mu$ and a standard deviation $\sigma$. The input is a cubic lattice (small dots) labeled by atomic species on BCC lattice sites (filled circle). Each atom species is placed into a separate channel. The illustration is for a four-component Mo-Nb-Ta-W system with four channels corresponding to Mo, Nb, Ta, and W (from top to down). The output is a reconstructed sample from the latent variable $z$. Fig. \ref{fig:HEA-order-parameter}(b) shows the 2D t-SNE plot of VAE 12D embedding on the test data for MoNbTaW. Various phases (colored by temperatures) are clearly separated in the 2D space. This indicates an order parameter can be defined in this space based on a distance metric. Such an order parameter is defined using the Manhattan distance, i.e., $Z^{op}$ (red, Eq.~\ref{eq:z}),
\begin{equation}
    \langle Z^{op}\rangle_T = \frac{1}{M}\sum_{X_j \in T}^M \sum_i^d|z_i(X_j)|, \label{eq:z}
\end{equation}
%The VAE based order parameter can the specific heat $C_v$ (Eq.~\ref{eq:c}) at different temperatures.
The equation can be used to calculate the VAE-based order parameter at various temperatures.
The inserted snapshots are representative configurations at three different temperatures, showing the microstructural changes from (1) strong B2 order, (2) partial B2 order, to (3) A2 (BCC) as the temperature increases. The corresponding data points in the latent space are also marked on the $Z^{op}$ vs. $T$ curve.

As a physical requirement, the second-order moment of the order parameter that has a physical meaning of susceptibility should have a peak at the phase transition point. Similar to the classical susceptibility, the VAE susceptibility is calculated from $Z^{op}$,
\begin{equation}
    \chi(Z^{op}) = \bigg{(}\langle(Z^{op})^2\rangle_T - (\langle Z^{op}\rangle_T)^2 \bigg{)}\frac{N}{T}. \label{eq:x}
\end{equation}
The $\chi(Z^{op})$ for several HEAs are shown in Fig. \ref{fig:HEA-order-parameter}(c). We compare the VAE susceptibility with experimental data. As indicated by the shadow bands (with arrows indicating an estimated range from experiments), it indeed correctly signals the phase transition temperatures for the HEAs.
% \begin{equation}
%     C_v = \bigg{(}\langle(E)^2\rangle_T - (\langle E\rangle_T)^2 \bigg{)}\frac{N}{T}. \label{eq:c}
% \end{equation}

The performance of the VAE order parameter is demonstrated through direct comparison with the classic SRO and LRO parameters (Fig. \ref{fig:HEA-order-parameter}(d)). The LRO parameter for each element and the total order parameters $\eta$'s in terms of LRO is shown in the left panel; the right panel shows the SRO for each element pair and total order parameters in terms of SRO. The overall trend reflected by the classic order parameters, i.e., increase or decrease of the ordering degree, is well captured by the new order parameter. Interestingly, the VAE order parameter is highly consistent with the LRO parameter in the high-temperature region; in contrast, it is more consistent with SRO parameters in the low-temperature area. This demonstrates that the new order parameter is not simply a surrogate of the classic order parameters. Instead, it has its unique features that merit further investigations.

\subsection{Mechanical-property prediction}
Here we will discuss the applications of ML algorithms to predict the mechanical properties, as illustrated in Fig.~\ref{fig:Application_bulk}. The important cases are summarized in Tab.~\ref{tab:mechanical}.

\begin{table}[!ht] 
%\footnotesize
\centering
\caption{A summary of the machine-learning applications for predicting the mechanical properties of alloys.}
\label{tab:mechanical}
\begin{tabular} {p{0.18\textwidth}p{0.18\textwidth}p{0.3\textwidth}p{0.05\textwidth} p{0.05\textwidth}}%{lllcccc }
\hline
\hline
%\hline
%\hline 
Methods & Target & Performance & Refs & Year \\ 
\hline
\hline
GP classification & Ductility enhancing elements & 92 \% consistency with the YSI descriptor for 76 chemical elements  &\hl{\cite{pei2019machine}} & 2019   \\ \hline
CCA+GA & Hardness (HV) & 5/7 of the synthesized HEAs within the 90\% prediction interval, 82 training data  & \cite{NatureCommCCAHard} & 2019\\ \hline
Linear and polynomial regression, SVM with different kernels,  CART, NN, KNN & Hardness (HV)  & RMSE of about 50 HV, 155 measured hardness data & \cite{WEN2019109} & 2019\\ \hline
Linear and Bayesian ridge regression, RF, KNN, and SVM & Yield strength & $R^2>0.95$ for RF, up to 44 data for each temperature subsets  &\cite{peng2020coupling}  & 2020 \\ \hline
SVM & Hardness & Five descriptors are identified as key features to the hardness; Recommended HEAs are synthesized and demonstrated very high hardness & \cite{YANG2022117431} & 2022 \\
\hline
Deep sets & Elastic Tensors & ML models trained from elastic property dataset of 7086 cubic quaternary HEAs, calculated with EMTO-CPA & \cite{Zhang2022-ia} & 2022 \\
\hline
RF & Yield strength of MoNbTaTiW and HfMoNbTaTiZr & 2.5\%  to 7.7 \% accuracy, 240 data & \cite{BHANDARI2021101871} & 2021 \\
\hline 
NN & $\gamma'$ phase volume fraction and yield strength & For $\mathrm{Ni_{32}Co_{28}Fe_{28}Cr_{3}Al_{3}Ti_{6}}$ HEA, ML prediction (53\% volume fraction and yield strength of 1.067 GPa) agrees with  measurements (50.4\% volume fraction and yield strength of 1.03 GPa) & \cite{ZHENG2021156} & 2021\\  \hline
Gradient-boosted decision trees & Elastic constants & For  $\mathrm{Al_{0.3}CoCrFeNi}$, 6\% error for bulk modulus, 10\% error for shear modulus, trained on 6,826 data & \cite{KIM2019124} &2019\\ \hline
\hline
\end{tabular}
\\ \raggedright
Acronyms in the table: Gaussian process (GP),  canonical correlation analysis (CCA), SVM (support vector machine), classification and regression tree (CART), neural network (NN), k-nearest neighbor (KNN), random forest (RF); yttrium similarity index (YSI), root-mean-square error (RMSE).
\end{table}

% Examples:
% \begin{itemize}
% \item  Use ML algorithms to predict the elastic constants of HEAs \cite{KIM2019124, DAI2020168}.

% \item Use ML algorithms to predict the strength, ductility, or hardness of HEAs \cite{NatureCommCCAHard,WEN2019109,peng2020coupling,BHANDARI2021101871,ZHENG2021156}.
% \end{itemize}

% We will use the following sub-sections to elaborate on these points.
% \subsection{Elastic constants}
% {\color{red} I remember the Berkeley people of materials project and their collaborators, associated students/postodc have done quite some work on this topic. Could you please kindly check them out and put the literature here? I can help with the writing if needed.}

% Wei Chen (illinois institute of technology); Peter K. Liaw (UTK), Science Adavances 2020/2021?

% Liang Qi, and his student, Acta Mater. 2021, GSFE in BCC. 

%\subsection{Strength, ductility and hardness}
Identifying reliable descriptors for mechanical properties by researchers is usually time-consuming since this may involve understanding the deformation mechanisms. In contrast, ML algorithms can identify the descriptors efficiently, given sufficient data from either experiments or simulations. Actually, if the data volume is substantial, ML models can even take all data, and the descriptor selection process can be ignored.
%Starting from the easily accessible experimental properties, ML can predict the mechanical properties of alloys. 
Probably due to the limited available experimental data as the training data for ML, there are few representative examples for HEAs, particularly the ML models for ductility. Here we use a binary Mg alloy as the example \hl{\cite{pei2019machine}}. The method shall be straightforward to extend to multicomponent Mg alloys \cite{pei2015rapid} and HEAs.

There are two widely acknowledged mechanisms to explain the enhanced ductility in Mg alloys upon alloying with rare-earth elements \hl{\cite{pei2020relation,sandlobes2012relation}}\cite{wu2018mechanistic}. One is based on the nucleation mechanism of the non-basal dislocations, whose critical role is guaranteed by the von Mises criterion \hl{\cite{sandlobes2012relation}}. The other explanation focuses on the mobility of the non-basal dislocation \cite{wu2018mechanistic}. Interestingly, both mechanisms can explain well available experimental observations. Moreover, the two descriptors associated with these mechanisms also successfully predict alloys with enhanced ductility. The origin of this correlated effect is revealed by a theoretical model \hl{\cite{pei2020relation}}. Therefore, it is of great significance if ML can reproduce the predictions of the two descriptors. This will demonstrate ML is not just an efficient method but also a reliable and accurate one. 

\begin{figure} %[h]
    \centering
    \includegraphics[width=0.9\linewidth]{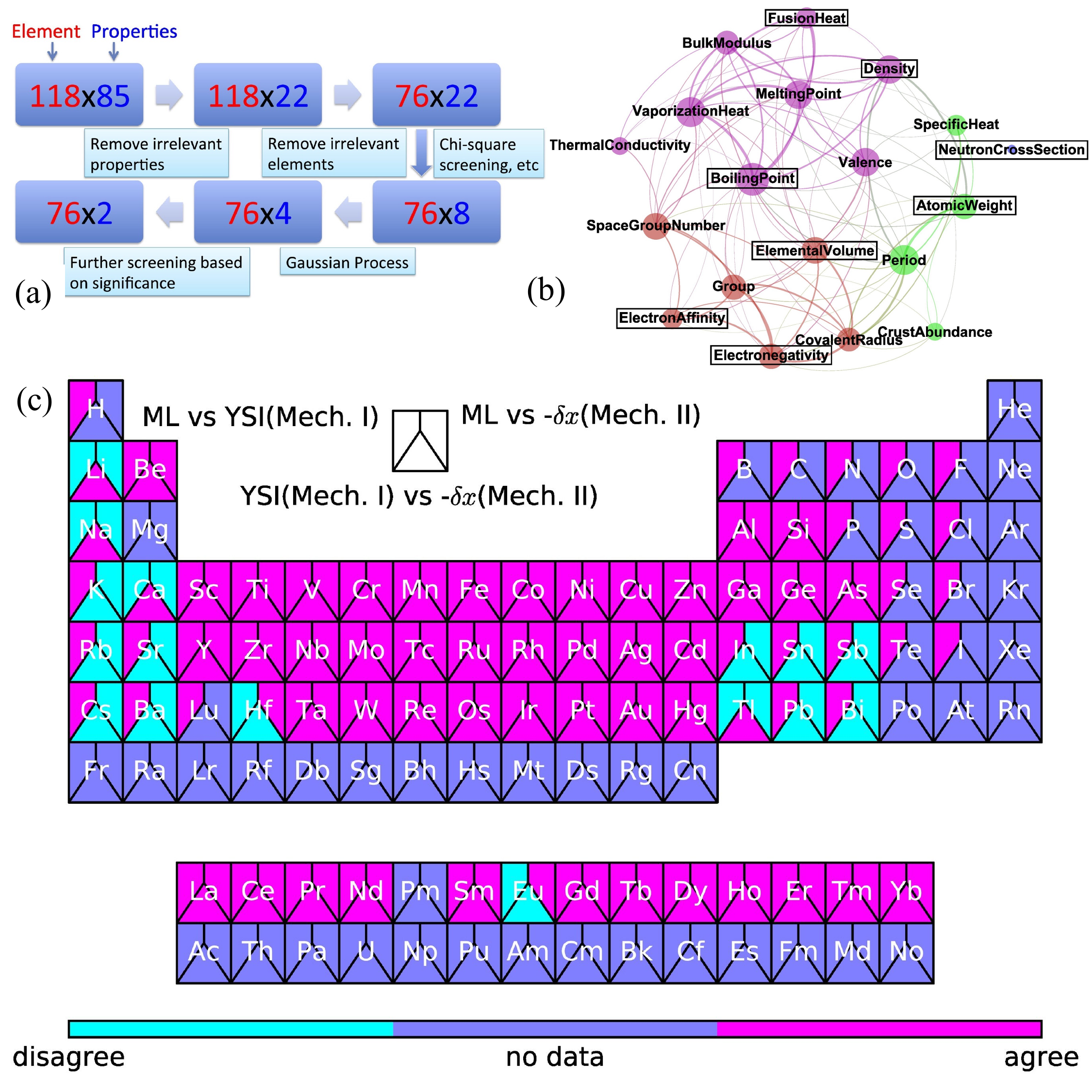}
    \caption{The machine-learning informed prediction of mechanical properties. Here the ductility of Mg alloys is taken as an example. (a) The whole process of our data analysis. (b) The correlation graph for the relevant properties in feature selection. An edge is formed between 2 nodes if Pearson's correlation between the two properties has a coefficient of $>$0.2 and edge thickness is weighted by the correlation value. Different colors represent different modularity classes based on each node's degree, and each node's size is scaled by its degree. The screening process identifies the boxed properties. (c) Comparison of the machine learning (ML) solution, YSI descriptor (Mechanism I), and $-\delta_X$ (Mechanism II). The cyan blocks represent where the two predictions disagree, the blue ones represent that no data is available, and the pink ones represent where the two predictions agree. The figures are reproduced from Ref. \hl{\cite{pei2019machine}} with changes.}
    \label{fig:MD2019}
\end{figure}

We performed DFT screening on 21 hexagonal close-packed elements, and 5 of those Mg alloys were synthesized and tensile tested \hl{\cite{sandlobes2014ductility}}. The experiments indeed confirmed the enhanced ductility in these alloys predicted by DFT. Given the insufficient experimental data for training, we adopt the Gaussian Process Classification (GPC) algorithm that can avoid overfitting and is particularly suitable for small datasets. The GPC model takes the 21 DFT predictions as the labels. To build an unbiased model, we consider all available 85 elemental properties and brutal-force down-select the ones that are most relevant to the ductility (Figs. \ref{fig:MD2019}(a)-(b)). After fine-tuning the model parameters, we show the ML predictions are highly consistent with the two descriptors \hl{\cite{pei2019machine}}. As shown in Fig. \ref{fig:MD2019}(c), the primary area is pink, indicating the consistency of ML results with the two descriptors/mechanisms.

Reliable ML models with carefully selected ML algorithms are not just mathematical black boxes but can also contribute to understanding physics in materials science problems. For example, when the two descriptors are not consistent in several predictions, ML as a third method can help pick the more promising descriptor whose prediction is consistent with it. The mechanism whose descriptors are in better agreement with ML is more likely the one active during deformation. Also, we can use ML to discover new descriptors that are able to reproduce the ML results, as descriptors usually do not need computer programming and are simpler than ML in applications. We give one example in Ref. \hl{\cite{pei2019machine}} and show it is possible to rediscover the descriptor for the mechanism I.
These results were summarized in Ref. \hl{\cite{pei2021mechanisms}} for more details.
%First-principles and machine learning predictions of elasticity in severely lattice-distorted high-entropy alloys with experimental validation : 

In addition to the above example, we briefly summarize several representative studies for other mechanical properties. Hardness is easy to measure and offers a quick means to estimate the mechanical properties of alloys. Given this fact, it is the targeted property in several ML studies for HEAs. 
Rickmann \textit{et al.} \cite{NatureCommCCAHard} proposed to use the canonical correlation analysis (CCA) and genetic algorithm to identify important features that contribute to the hardness of HEAs. As a result, they identified promising HEAs and confirmed their exceptional hardness in experiments.
In a similar study, Wen \textit{et al.} designed HEAs with high hardness assisted by ML models \cite{WEN2019109}. First, they constructed several ML models to estimate the hardness, including a linear regression model, a polynomial regression model, support vector regression models with different kernels (linear, polynomial, and a radial basis function), a regression tree model (CART), a neural network model and a k-nearest neighbor model. Then, guided by the models, they synthesized a few alloys with a hardness 10\% higher than the best alloy in the training dataset.

Peng \textit{et al.} successfully predicted the yield strength for high-temperature multicomponent steels using ML models \cite{peng2020coupling}. Adopting the actual processing conditions and information from CALPHAD, the researchers developed ML models that are able to reproduce the existing experimental yield stresses. Using the random forest regressor, Bhandari {\it et al.} \cite{BHANDARI2021101871} predicted the yield stresses of MoNbTaTiW and HfMoNbTaTiZr at 800 $^{\circ}$C, 1200 $^{\circ}$C, and 1500 $^{\circ}$C. The results were compared with experiments, and good agreement was obtained. Moreover, Zheng {\it et al.} \cite{ZHENG2021156} fed the data of compositions-microstructure-properties into an artificial neural network to predict the $\gamma'$ phase volume and yield stress. Guided by the ML model, they explored the relevant compositional space and designed a HEA with superior strength and ductility. The experiment validated the excellent mechanical property of the HEA, which gives a yield strength of 1.31 GPa and a tensile elongation of 15\%.

Except for hardness and yield stress, there are also studies focusing on the elastic constants, which are directly calculated by DFT. For example, Kim \textit{et al.} carried out detailed calculations of the elastic constants of an FCC HEA AlCoCrFeNi, focusing on the effect of lattice distortion \cite{KIM2019124}. They found that including the lattice distortion in DFT brings the elastic constants closer to experiments. Moreover, the ML model trained on a large dataset of inorganic structures also accurately predicts the HEA's elastic properties.

\section{Challenges and Outlook}
\subsection{Challenges}
 Despite ML's advantages for HEAs, challenges still need to be addressed to realize their potential fully. The first is model confidence, i.e., understanding when the model gives accurate prediction and when it may not. Without such knowledge, it would be very risky to trust the prediction of an ML model, as illustrated in some well-known ``adversarial attack'' examples \cite{8601309}. One standard tool for this problem is uncertainty quantification (UQ), which attempts to estimate the model uncertainty via statistical methods. It is also desirable to carry out an objective, and extensive benchmark of various ML potential and models \cite{ Hart_NPJ_2019,Ong_JPC_2020,  Hart_NPJ_2021, Benchmark_GNN} \hl{\cite{lavin2021simulation}}, similar to the field of DFT \cite{Lejaeghereaad3000}. So, some heuristic rules can be established on the model's performance and applicability. 

 Another challenge is to learn the underlying physics efficiently. As previously emphasized, the arrangement of atoms in materials has various symmetries, such as permutation, translation, rotation, and crystal symmetries, and the movement of atoms follows physical laws. Therefore, only specific points and lines represent realistic systems in the high-dimensional feature space. Currently, the most popular approach is to use hand-crafted symmetry-preserving descriptors, as introduced in section \ref{symmetry_descriptor}. Despite the considerable success of these methods, an end-to-end scheme that automatically captures physics-relevant descriptors on the 3D lattice is still highly attractive due to its broader applicability. One such example is a graph neural network (GNN), where the local neighboring atoms and bonds are represented by the nodes and edges of a graph rather than a set of pre-calculated descriptors. However, end-to-end models generally demand a larger dataset to train due to their enhanced flexibility. For example, Ref.\cite{Benchmark_GNN} demonstrates that a series of GNN methods require one or two magnitudes larger datasets to gain similar accuracy as descriptor methods. Therefore, addressing this challenge requires further progress in both ML models and data sampling.

The third challenge is obtaining a high-quality dataset, which is difficult for a few reasons. First, experimental and simulation data face the problem that different research groups usually make measurements under other conditions. Combining these data requires careful calibration. Moreover, both experimental and simulation data are time-consuming to obtain. As a result, the dataset size is generally small, as demonstrated in Tab. \ref{tab:thermo}. The small-data problem is even more severe for HEAs than other materials, considering their vast design spaces. A series of methods to improve the dataset has been discussed in section \ref{Data_Aug}. However, these methods are generally still developing and cannot completely solve the small-data problem. Considering the central role of datasets for ML models and the increase in model complexity, constructing high-quality datasets will be essential for the development of ML for HEAs in the foreseeable future. 

Datasets play a central role in the quality of ML models. This role will also profoundly affects sharing of experimental data among the materials science community. For example, in scientific literature, it is common only to report materials that demonstrate the desired properties, while the failed trials are ignored due to the assumption that they are of little interest to other researchers. However, such an assumption is not valid for ML methods. The negative results are also valuable for quality-critical data balance, as exemplified by a study on inorganic-organic hybrid materials \cite{Raccuglia2016-ta}. Therefore, to facilitate the development of ML models, researchers should be encouraged to publish negative results along with positive ones. Another critical point is that, in many cases, only relatively small data is available. This fact is common for many physical properties due to the cost of experimental measurements. For small datasets, ML models, such as linear regression and support vector machines, are expected to perform better than the more complicated ones, such as deep neural networks, which can easily lead to overfitting. Finally, it is also essential to report the measurement details to facilitate the use of physical-property data in ML. This is because the measured values of alloys' properties significantly depend on conditions such as the samples' synthesis, heat treatment, and measurement methods. Therefore, before feeding the measured values to ML models, it is vital to carry out careful data cleaning and calibration to ensure that these data are on the same footing. Without such a procedure, the quality of the property dataset can be compromised, which strongly affects the performance of the ML models.
%}
%\subsection{Knowledge mining}
%For knowledge mining, the natural language processing (NLP) methods can be employed to analyze the abundant scientific publications in materials science. One commonly used NLP technique is Word2vec \cite{mikolov2013efficient,mikolov2013distributed},  which is an unsupervised word-embedding method that use a two-layer neural network to represent the contexts of words. The training is accomplished by requiring the output vector $w_o$ maximizing the log likelihood with respect to the training dataset $(w_i: w_c)$, where $w_i$ are the input vectors and $w_c$ are the contexts, or nearby words of $w_i$. After training, words with similar contexts would generally have similar representation in the Word2vec embedding. Using Word2vec, V. Tshitoyan \textit{et al.}  \cite{Tshitoyan2019} demonstrated that chemical knowledge can be extracted automatically by unsupervised learning from scientific literature, which facilitates the search of thermoelectric materials. Another method to learn from materials science publications is knowledge graph, which can be employed for applications such as building databases of derived properties, finding unknown physical correlations, and identify research trends \cite{Krenn201914370}. For example, D. Mrdjenovich \textit{et al.} \cite{MRDJENOVICH2020464} constructed the ``propnet'' knowledge graph, which significantly expands existing datasets of material properties by using graph traversal algorithms to find and calculate related properties. 

\subsection{Uncertainty quantification}
\begin{figure} %[h]
    \centering
    \includegraphics[width=0.98\linewidth]{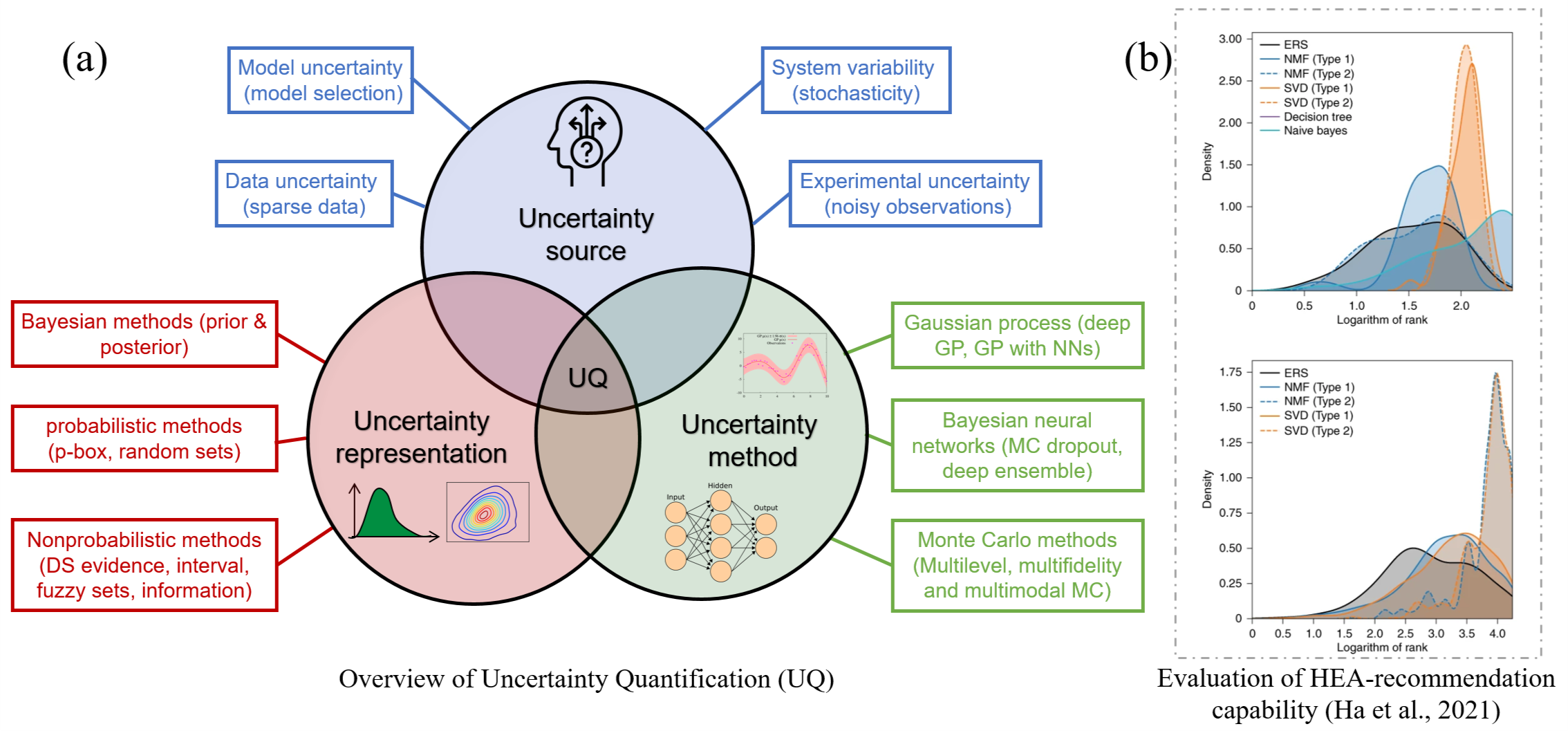}
    \caption{Uncertainty quantification (UQ). (a) The components of UQ in ML models: uncertainty sources, uncertainty representations, and uncertainty methods. Each component is comprised of multiple methods. (b) One typical example for the UQ applications in HEA design by Ha {\it et al.} \cite{ha2021evidence} after adjustments. They proposed an evidence-based recommender system for HEAs based on the Dempster-Shafer theory, which is a general framework for reasoning with uncertainty. The two sub-figures show the probability density functions for the HEA rankings in different test sets. The model performance is better when the probability density is distributed toward 0.}
   \label{fig:UQ}
\end{figure}

Most of the time, ML algorithms are used as ``black-box'' methods; therefore, it is essential to conduct UQ to determine the reliability of the predictions \cite{ghanem2017handbook}.
The confidence of ML models merits particular attention for applications in HEAs, where the dataset size is typically not large, but the design spaces are huge \cite{george2019high}. One critical approach to overcome this challenge is uncertainty quantification (UQ), which gives an estimate of the uncertainty for each prediction in a rigorous way \cite{tran2020methods}.
In practice, there are different approaches to estimating the uncertainty of models \hl{\cite{bi2021towards}}, as illustrated in Fig.~\ref{fig:UQ}. For example, Ha {\it et al.} \cite{ha2021evidence} recently developed an evidence-based material recommender system (ERS) for HEAs, which utilizes Dempster-Shafer (DS) theory, a general framework for reasoning with uncertainty. First, the ERS system rationally considers each dataset as a source of evidence. Then, it combines the evidence to draw the final recommendation of HEAs, where the uncertainty can be quantitatively assessed. 
Except for DS evidence theory, there are also many other UQ methods, including fuzzy sets \cite{zadeh1996fuzzy}, interval methods \cite{moore1979methods}, and information theory \cite{klir2006uncertainty}, as well as probabilistic approaches such as Bayesian methods \cite{raftery1995bayesian} and probability boxes \cite{ferson2015constructing}. For example, Zhang and Shields \hl{\cite{zhang2018quantification}} proposed a unified Bayesian framework that allows quantifying uncertainties resulting from small datasets using multimodel inference and efficiently propagating the uncertainties through optimal importance sampling. They further investigated the effect of prior probabilities on the evaluation of uncertainties given limited datasets. Moreover, sampling-based Monte-Carlo methods can also be employed for UQ. These methods are straightforward to implement but typically computationally intensive. Several modern Monte-Carlo methods \cite{zhang2020modern}, including multilevel Monte-Carlo \cite{heinrich2001multilevel,giles2008multilevel,giles2015multilevel}, multi-fidelity Monte-Carlo \cite{peherstorfer2016optimal,peherstorfer2018survey} and multimodel Monte-Carlo methods \hl{\cite{zhang2018quantification,zhang2018uncertainty,zhang2019efficient,zhang2020quantification}}, are developed to address the shortcomings. A confidence interval on the model prediction can be drawn by identifying and quantifying the sources of the uncertainty. 
This not only allows users to understand the prediction reliability but also facilitates the implementation of active learning \hl{\cite{zhang2021scalable,zhang2021self}} by adding data in the region of high uncertainty, which better fills the design space \hl{\cite{shields2016generalization}}.  

%\subsection{Microscopic image analysis}
\subsection{End-to-end (automatically learned) descriptors}
A different class of methods that do not explicitly depend on the atomic energy decomposition in Eq.~\ref{eq:atomic_energy} are convolutional neural network (CNN) based models, such as the graph convolutional neural networks (GCNN) \cite{pmlr-v70-gilmer17a, PhysRevLett.120.145301, doi:10.1063/1.5019779} and three-dimensional (3D) CNN \cite{10.1145/3042064, CECEN201876}. In principle, CNN models can find the important features automatically compared to descriptor-based models. The atomic environment is learned with convolution filters to determine the total energy. CNN models usually use a large number of parameters and thus are, in general, more complex than descriptor models. Therefore, it requires more training data \cite{Benchmark_GNN}. The most popular CNN model for atomic interactions is GCNN, which demonstrates excellent interpretability \cite{PhysRevLett.120.145301, 8954227} by maintaining the concepts of atoms and chemical bonds. Compared to conventional CNN, GCNN employs graph-structured data rather than 2D or 3D grids to represent the arrangement of atoms in the system. In each convolutional layer, the ``atom'' nodes update their values by convoluting with their surrounding atoms and bonds and output a graph with the same size and structure as the input graph. This process is very similar to the numerical renormalization group. After a few convolutional layers, the feature vector represents the ``renormalized'' atoms. The convolutional layers are then followed by pooling and fully connected (FC) layers to determine the total energy. By replacing these pooling and FC layers with a simple summation, it is easy to see that the outputs of the last convolutional layer can be interpreted as local atomic energies. Different GCNN models have been proposed, such as SchNet \cite{10.5555/3294771.3294866},
MEGNet \cite{ChenGN}, DimeNet \cite{gasteiger_dimenet_2020}, and GemNet \cite{gasteiger_dimenet_2020}, and their implementation can be found in the Pytorch geometric package \cite{https://doi.org/10.48550/arxiv.1903.02428}. A benchmark of common GCNN models in various bulk, 2D, and nano-structure materials are presented by Fung et al. \cite{Benchmark_GNN}. GCNN models have been widely used for predicting the physical properties of both molecules \cite{10.1093/bib/bbz042, ijms20143389, 10.1093/bib/bbz042} and crystals \cite{PhysRevLett.120.145301, PhysRevMaterials.4.063801, Ong_GCNN}.
However, the adoption of GCNN for the atomistic simulations of complex materials, particularly HEAs, has been rare. One reason is that while end-to-end models generally have better representation capability, they are also prone to overfitting and require a much more extensive training dataset to achieve high prediction accuracy.

%By combining a high-fidelity model (HFM) with inexpensive but nonetheless less accurate low-fidelity models (LFM), i.e., the so-called multi-fidelity model, can be designed to work well for both large and small datasets.

\subsection{Machine-learning guided inverse materials design}

Materials design is central to materials science, which can also benefit from the introduction of ML techniques. The design of new materials requires identifying the possible element mix to produce the targeted material performance. In contrast to researching the mechanisms underlying the properties of known materials, this is essentially an inverse design process \cite{sanchez2018inverse,zunger2018inverse,tarantola2006popper,yu2013inverse} \hl{\cite{zhang2021efficient,fung2021inverse,Pei2021AS,fung2022atomic}}. The targeted material properties are the consequences of many factors, including the alloy components and the processing. Given this reduction feature, the same properties can result from many different combinations of the factors. Due to this asymmetry of information flow and the lack of constraints, materials design is indeed a very challenging task, accompanied by many failures.

Edison's trial-and-error method has long been a norm for the design of new alloys for thousands of years. However, this method has been confronted with the challenge of increasingly more components and the resulting complicated hierarchical microstructure. Fortunately, we have accumulated many experimental data that can be used in ML models to design complex alloys. Moreover, new experimental methods and algorithms armed with modern, powerful supercomputers result in theoretical databases to overcome the common problem of insufficient data for ML models. We provide an example of using microscopy images in the inverse design of complex alloys. Given the lack of typical examples of high-entropy alloys, we adopt a model for steels with about 20 components. Albeit with less concentrated elements, the many components of the steels indeed pose challenges as serious as high-entropy alloys, if not more severe. The same method is applicable for high-entropy alloys.

\begin{figure} %[h]
    \centering
    \includegraphics[width=0.9\linewidth]{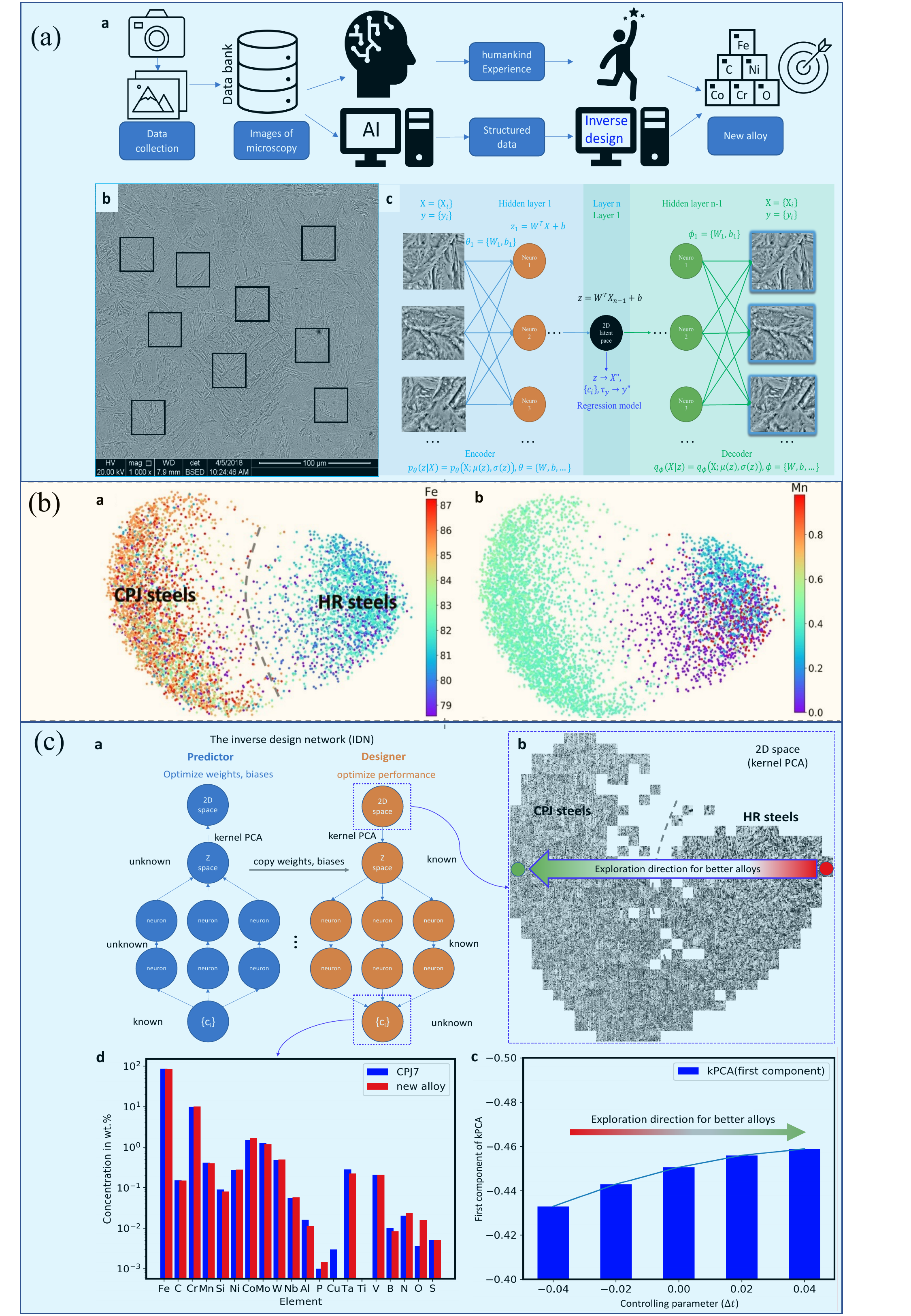}
    \caption{The machine-learning informed prediction of mechanical properties and inverse material design. (a) The neural-network model constructed based on the variational autoencoder (VAE) model and the pre-processing of the training data. (b) The kernel principal component analysis (kPCA) of the latent space of the neural network model. (c) The inverse material designer based on the regression branch of the model in (a). The figures are reproduced from Ref. \hl{\cite{pei2019machine}} with adjustments.}
    \label{fig:AS2021}
\end{figure}

We provided a neural network solution for the inverse material design based on the knowledge obtained from microstructure images (Fig. \ref{fig:AS2021}a). The inputs are the microstructure images and mechanical properties measured experimentally. The images are collected using a scanning electron microscope (SEM) with some post-processing. The training data is fed into a neural-network model that consists of three sub-models, i.e., the encoder, the decoder, and a regression model. Each of them has its unique function. Details are referred to Ref. \hl{\cite{Pei2021AS}}. As one of the key results, two examples using Fe and Mn concentrations as labels are shown in Fig. \ref{fig:AS2021}(b). The model can separate the steels with different components and mechanical properties (here, yield stresses at room temperature). 

The performance of the model to differentiate the complex steels encourage us to go one step further for the inverse design (Fig. \ref{fig:AS2021}(c)). The main focus here is to show the main ideas. The regression model was originally a predictor that predicts material properties based on alloy components and compares them with the measured ones to optimize the model parameters. We copy the structure and optimized parameters of the regression model to construct a designer. The designer has a reversed information flow, albeit with the same structure and model parameter. It allows us to explore and determine the optimal microstructure of the alloys and then transform the microstructure into alloy components. This actually realizes the challenging task of inverse design.
The inverse design method proposed here can also be readily applied to other material systems, given sufficient microstructure images. It is expected that many similar methods will emerge and provide new momentum for designing new materials with lower costs and shorter periods.

Another vital application of ML in materials science is the automatic analysis of scientific images from technologies such as X-ray diffraction (XRD), transmission electron microscope (TEM), scanning electron microscope (SEM), and atom probe tomography (APT). The primary motivation for adopting ML models rather than doing it by researchers is to enhance the analysis efficiency and avoid possible artificial errors \cite{decost2017exploring}. For example, Li {\it et al.} \cite{CNN_AlLiMg} employed CNN to identify the L12-type ordered nanostructures in the APT data of an FCC Al-Li-Mg system. This task is prohibitively tricky for researchers due to the millions of data points. In another example, Kaufmann {\it et al.} \cite{Kaufmann564} classified images with two commonly used neural network architectures, i.e., ResNet50 and Xception, for the autonomous identification of crystal structures based on electron backscatter diffraction (EBSD) patterns. Each model has an overall accuracy of >90\%. Currently, this research direction is still in its infancy. Ragone {\it et al.} employed a fully convolutional neural network (FCNN) to determine atomic positions from the pixels' intensities of STEM images. As a result, they obtained a sufficiently high level of confidence \cite{RAGONE2022110905}. Finally, data augmentation techniques are employed by Ma {\it et al.} \cite{USTB_DataAug} to mitigate the small data size problem by generating synthetic data of polycrystal images. 
With the further improvement of the methods and the accumulation of experimental data, we can expect that autonomous image analysis can be used to build practically useful models that directly predict the mechanical properties from the microstructure images with high accuracy.

\section{Conclusions}
The synergy of ML and materials science promotes the formation of a burgeoning and exciting field with exponential growth. High-entropy alloys are particularly interesting candidates to demonstrate the power of ML due to their superior mechanical properties, vast compositional space, and complex chemical interactions. The ML studies of high-entropy alloys can be grouped into different categories depending on the specific motivations. On the one hand, based on DFT datasets, ML methods can be used to construct atomic interaction models (AIMs) to describe the complex interactions in multicomponent alloys. The AIMs help to avoid the time-consuming first-principles thermodynamics simulations. On the other hand, ML can also be used to build models to predict bulk properties such as phase formations, crystal structures,  elastic constants, and yield strengths, starting from high-throughput experimental or calculation data.

Atomistic simulations require high-fidelity predictions of atomic energies and forces; therefore, retaining the underlying symmetry of atomic systems is crucial. To this end, a series of ML AIMs have been proposed \cite{Ong_JPC_2020, Asta_NC_MTP_2021}, including HDNNP, GAP, SNAP, MTP, LRP, Bayesian CE. These ML models have enabled simulations with near DFT accuracy at length and time scales inaccessible with conventional methods \cite{Nature_2021_Silicon, 10.5555/3433701.3433707}. The excellent combination of efficiency and accuracy benefits HEAs, where large-scale simulations are required to understand the origins of the exceptional mechanical properties in HEAs. Indeed, ML AIMs have been successfully applied to investigate the thermodynamics and mechanical properties of HEAs via MC or MD simulations, as summarized in Tab.~\ref{tab:t2}. 
For bulk property models (BPMs), various ML methods, such as NN, SVM, GP, and CART, are employed. The datasets can be obtained from experiments or theoretical calculations. 
One crucial procedure for building BPMs is the selection of features and models. Techniques such as feature analysis, model regularization, and out-of-sample validation are needed. The huge compositional space and complex chemical interactions of HEAs render ML models advantageous over empirical rules. ML models have demonstrated excellent performance in predicting the bulk properties of HEAs, such as phase formations and mechanical properties, as summarized in Tab.~\ref{tab:thermo}. Moreover, the ML models have been applied to guide the design of HEAs, with exceptional mechanical properties validated by experimental measurements \cite{ML_Phase_sensitivity, ZHENG2021156}. 

Finally, some promising research directions are discussed and illustrated with examples, including uncertainty quantification of ML models, end-to-end descriptors, and the ML-assisted inverse design of materials. These research directions have attracted some researchers and merit more research efforts due to their importance. Employing ML methods to study high-entropy alloys is expected to continue expanding rapidly, and many opportunities await further exploration. 

\section{Acknowledgements}
The work of X.L. and Z.P. was supported by the U.S. Department of Energy, Office of Science, Basic Energy Sciences, Materials Science and Engineering Division. The work of J.Z. was supported by the U.S. Department of Energy, Office of Science, Office of Advanced Scientific Computing Research, and the Artificial Intelligence Initiative at the Oak Ridge National Laboratory (ORNL).
\bibliographystyle{unsrt}  
\bibliography{refers.bib}

\end{document}